\documentclass[12pt,preprint, epsfig]{aastex}

\shorttitle{Not-So-Simple Stellar Populations in NGC 1831 and NGC
  1868} 
\shortauthors{Chengyuan Li, Richard de Grijs, and Licai Deng}
\usepackage{color}

\begin{document}

\title{Not-So-Simple Stellar Populations in the Intermediate-age Large
  Magellanic Cloud Star Clusters NGC 1831 and NGC 1868}

\author{
Chengyuan Li,\altaffilmark{1,2}
Richard de Grijs,\altaffilmark{1} and
Licai Deng\altaffilmark{2}
}

\altaffiltext{1} {Kavli Institute for Astronomy \& Astrophysics and
  Department of Astronomy, Peking University, Yi He Yuan Lu 5, Hai
  Dian District, Beijing 100871, China; joshuali@pku.edu.cn,
  grijs@pku.edu.cn}
\altaffiltext{2} {Key Laboratory for Optical Astronomy, National
  Astronomical Observatories, Chinese Academy of Sciences, 20A Datun
  Road, Chaoyang District, Beijing 100012, China}

\begin{abstract}
Using a combination of high-resolution {\sl Hubble Space
  Telescope}/WFPC2 observations, we explore the physical properties of
the stellar populations in two intermediate-age star clusters in the
Large Magellanic Cloud, NGC 1831 and NGC 1868, based on their
color--magnitude diagrams. We show that both clusters exhibit extended
main-sequence turn-offs. To explain the observations, we consider
variations in helium abundance, binarity, age dispersions, and
fast rotation of the clusters' member stars. The observed narrow main
sequence excludes significant variations in helium abundance in both
clusters. We first establish the clusters' main-sequence binary
  fractions using the bulk of the clusters' main-sequence stellar
  populations $\ga 1$ mag below their turn-offs. The extent of the
turn-off regions in color--magnitude space, corrected for the
  effects of binarity, implies that age spreads of order 300 Myr may
be inferred for both clusters if the stellar distributions in
  color--magnitude space were entirely due to the presence of multiple
  populations characterized by an age range. Invoking rapid rotation
of the population of cluster members characterized by a single age
also allows us to match the observed data in detail. However, when
taking into account the extent of the red clump in color--magnitude
space, we encounter an apparent conflict for NGC 1831 between the age
dispersion derived from that based on the extent of the main-sequence
turn-off and that implied by the compact red clump. We therefore
conclude that, for this cluster, variations in stellar rotation rate
are preferred over an age dispersion. For NGC 1868, both models
perform equally well.
\end{abstract}


\section{Introduction}

The majority of stars in a given star cluster all formed in close
proximity to each other at roughly the same time. They are thought to
have originated from the same progenitor molecular cloud and, hence,
they are characterized by the same chemical composition. This idea is
the basis of the common notion that all observed stars in such a
cluster should have roughly the same age and metallicity, which is in
essence equivalent to saying that the cluster is composed of a single
(simple) stellar population (SSP). Although the SSP approximation
remains valid for the bulk of the stellar populations in most observed
star clusters, the discovery of extended main-sequence (MS)
  turn-offs (TOs) and multiple stellar populations in many globular
clusters (GCs), as well as in massive extragalactic clusters of any
age, challenges this simple picture. Based on the SSP assumption, one
expects observed clusters to display narrow sequences in their
color--magnitude diagrams (CMDs), including as regards their MS,
subgiant branch (SGB), red-giant branch (RGB), a compact
horizontal-branch (HB) clump, and a well-defined, narrow MS
TO. However, counterexamples have been found that challenge these
expectations for almost all of these narrow and tight CMD
features. Some clusters -- including NGC 419 \citep{Glat08a}, NGC
1751, NGC 1806 \citep{Milo09}, NGC 1783 \citep{Mack08}, NGC 1846
\citep{Mack07,Milo09}, and NGC 2209 \citep{Kell11} -- exhibit extended
or dual TOs. Some old GCs also display double or multiple MSs, such as
NGC 2808 \citep{Piot07} and NGC 6397 \citep{Milo12}. \cite{Milo08}
found that the GC NGC 1851 is characterized by two distinct SGBs. Some
GCs, such as $\omega$ Centauri \citep{Piot05,Soll07}, NGC 288
\citep{Piot13}, and M22 \citep{Lee09}, exhibit double or multiple MSs,
SGBs, and/or RGBs; Terzan 5 even shows clear, double HB clumps
\citep{Ferr09}.

Although some of these features are only found in specific example
clusters, extended or double MS TOs seem to occur rather
commonly. \cite{Milo09} suggested that this may also be an ordinary
feature of {\it intermediate-age} star clusters in the Large
Magellanic Cloud (LMC). They found that roughly 70\% of their
  sample of intermediate-age clusters display extended or double TOs.
Different models have been proposed to account for these observations,
 including those involving the presence of chemical
  inhomogeneities, internal age dispersions, rapid stellar rotation,
  and possible selection effects \citep[for a discussion,
    see][]{Kell11}. An initially promising explanation involved the
assumption of helium inhomogeneities \citep[for general overviews,
  see][]{Rood73,Rood89,Fusi97}. However, it seems that this model only
works properly for the multiple populations in ({\it old}) GCs, e.g.,
in NGC 2808 \citep{Anto05,Piot07} or $\omega$ Centauri
\citep{King12}. Since the helium-enrichment scenario predicts the
presence of a secondary MS at bluer colors than those of the main
stellar population, such models can be used to fit the observed color
bifurcation of GC MSs very well. However, intermediate-age clusters
{\it also} exhibit extended TOs, while their MSs remain narrow
\citep[see, e.g.,][]{Milo09}. As a consequence, few authors consider
helium-abundance differences in the context of the extended TOs seen
in the CMDs of intermediate-age clusters. In this paper, we will
confirm that assuming differences in the helium abundance indeed
cannot explain our observations of the CMDs of two intermediate-age
LMC clusters.

An alternative model, which assumes that the cluster stars have been
forming continuously, with an age dispersion of roughly 300 Myr, has
been more successful in explaining the extended TOs observed for
intermediate-age clusters. However, the origin of such an age
dispersion as incorporated in this model remains
unclear. \cite{Erco08} and \cite{Goud09} suggested that the ejecta of
first-generation asymptotic giant-branch (AGB) stars might be the main
contributors, but this again necessitates helium self-enrichment. Some
authors have also suggested that this postulated age dispersion may
originate from dynamical interactions, leading to differently aged
populations that are each characterized by a specific chemical
abundance. For instance, \cite{Mack07} proposed that mergers of two
star clusters with an age difference of $\geq 200$ Myr might be the
origin of the observed multiple populations in their sample
clusters. Similarly, \cite{Bekk09} proposed a new scenario involving
interactions or mergers of star clusters and star-forming giant
molecular clouds. Although it is still unclear why such dynamical
processes might have acted in a global manner and affected many
intermediate-age LMC clusters, the age-dispersion model has been
successful in reproducing most of the observations.

A recently proposed, competing scenario invokes fast stellar rotation,
the effects of which become particularly pronounced in MS TO
stars. This model can also reproduce the observations, while -- more
importantly -- it avoids challenging the SSP assumption. The main
principles of this latter model are that the centrifugal force
resulting from rapid rotation causes a star to expand, thus decreasing
its effective temperature. The reduced effective gravity also results
in a lower luminosity \citep{Bast09,Yang13}. \cite{Roye07}
investigated a large sample of B--F-type solar-neighborhood stars and
found that almost all of these stars are fast rotators. Although the
functional form of the distribution of stellar rotation velocities in
compact star clusters is still unclear, adopting the assumption that
MS TO stars are affected by rapid rotation seems reasonable (for a
detailed discussion, see Section 4). A number of authors have
investigated this issue. They all found that rapid rotation can
reproduce extended or double MS TOs \citep{Bast09,Li12,Yang13},
provided that one adopts a proper distribution of rotational
velocities.

Both of these suggested models have their own shortcomings. As regards
the age-dispersion model, if the observed extended-TO stars originate
from AGB ejecta, they cannot generate the abundance of observed
second-generation stars \citep{Bast09}, unless we assume that these
clusters would have been much more massive originally (i.e., by one or
two orders of magnitude) than they are today \citep{Bast13}. The
problem with the cluster--cluster merger scenario is that this process
does not seem to occur commonly in ordinary molecular clouds
\citep{Goud09}, and none of the young clusters studied in detail to
date display the requisite properties supporting such a scenario
\citep{Plat12}. However, as pointed out by \cite{Gira11}, it is still
possible that the homogeneity in the LMC clusters' age distribution
may be owing to dramatic large-scale events that may have happened in
the LMC's history. The fast-rotation model is currently debated in
relation to the predicted prolonged MS lifetime problem. \cite{Gira11}
claimed that rapid rotation will allow MS stars to live longer,
because the increased rotation speed allows more hydrogen to move to
the stellar core through convection. If we take this prolonged MS
lifetime into consideration, the cluster's resulting synthetic CMD
will be similar to the CMD of an SSP containing unresolved binary
systems (see Section 3). More recent studies have rendered this
argument weaker, however. \cite{Yang13} pointed out that
\cite{Gira11}'s conclusions are based on the assumption of efficient
mixing in stellar atmospheres. However, if we adopt a different mixing
efficiency, the fast-rotation model may still work \citep{Yang13}. In
addition, their results show that rapid rotation will work in
particular for intermediate-age clusters.

In this paper, we analyze the CMDs of the intermediate-age LMC
clusters NGC 1831 and NGC 1868. Both have similar ages and masses, and
contain comparable numbers of stars. The photometric observations on
which we base our results have been obtained with the {\sl Hubble
  Space Telescope} ({\sl HST}; see Section 2). Their CMDs exhibit
narrow MSs, clearly extended MS TOs, and compact red clumps (RCs). We
will show that both the age-dispersion model and rapid stellar
rotation can reproduce the observations. On the basis of an
exploration of the detailed distribution of MS TO stars in CMD space,
we obtain the best-fitting age distributions. Adopting the
distribution of rotation speeds of \cite{Roye07}, the observed
distributions of MS TO stars in the CMDs can also be reproduced in
detail. We further examine the constraint imposed by the compactness
of the RCs, and find that the age dispersion implied for NGC 1831
should be less than 100 Myr, in which case the fast-rotation model
becomes a more likely contributor to the broadened TO region
associated with this cluster. For NGC 1868, on the other hand, both
models work equally well.

This paper is organized as follows. The data-reduction approach is
outlined in Section 2. In Section 3, we first rule out the possibility
of different helium abundances as the driver of the occurrence of
secondary populations, and then show to what extent the age-dispersion
model reproduces the data, corrected for the effects caused by
  the presence of a significant fraction of MS binary systems. We
show that the age dispersion in NGC 1831 can be further constrained,
to within 100 Myr, by taking into account its compact RC. In Section
4, we discuss the performance of models involving rapid rotation. A
discussion and our conclusions are contained in Section 5.

\section{Data Reduction}
\subsection{CMD determination}

The data sets pertaining to NGC 1831 and NGC 1868 were obtained as
part of {\it HST} program GO-7307 (PI Gilmore), using the Wide-Field
and Planetary Camera-2 (WFPC2). These two clusters were originally
selected, because they form a pair in terms of their ages, they have
similar masses \citep{grijs02b}, and they are also located at similar
distances from the LMC's center \citep{West90,Bica96}. The latter
property minimizes any differences caused by the LMC's tidal field. As
we will see, their stellar (field) backgrounds are sparse and allow
for easy decontamination. Both clusters were observed through the
F555W and F814W filters, which roughly correspond to the
Johnson--Cousins $V$ and $I$ bands, respectively. We will henceforth
refer to these {\sl HST} filters as $V$ and $I$. In both filters,
three images with total exposure times of, respectively, 2935 s and
3460 s were obtained. Among these three images, two were taken with
the clusters' center regions located on the Planetary Camera (PC)
chip; the total exposure times of these images were 435 s in the $V$
band and 960 s in $I$ \citep[for more details about the data sets,
  see][]{grijs02b}. The longer-exposure images (characterized by
exposure times of 2500 s in both the $V$ and $I$ bands) were centered
on locations representative of the clusters' half-light radii. Images
with exposure times of 1200 s in the $V$ band and 800 s in $I$ cover a
nearby, representative field region for both of our clusters
\citep[see also][]{Kerb02}. These images are used to correct for
background contamination \citep[for more details of the relevant
  procedures used, see][]{grij02a,Hu10}.

We used the {\sc HSTphot} package \citep{Dolp00} to perform photometry
on the images. {\sc HSTphot} is a specialized photometry package for
analyzing {\sl HST}/WFPC2 images \citep{Dolp05,Hu10}, which can be
used to automatically deliver {\sl HST}/WFPC2 photometry and the
corresponding photometric uncertainties. We specifically chose to use
point-spread-function photometry. The photometric catalogs resulting
from all three images were combined, separately for both filters. The
stars in the combined catalog all have associated photometric
uncertainties of less than 0.22 mag. We applied the same photometric
procedures to our observations of the nearby field region \citep[for
  details, see][]{Li13}. By virtue of the long exposure time of
  the final, combined data set, the photometric uncertainties in the
  magnitude range of interest ($V \la 24$ mag; see below) are much
  smaller than the width of the clusters' MSs or the extent of their
  TO regions. At these magnitudes, our photometric completeness levels
  are $> 80$\% \cite[cf.][]{grijs02c}.

\begin{figure}[ht!]
\plotone{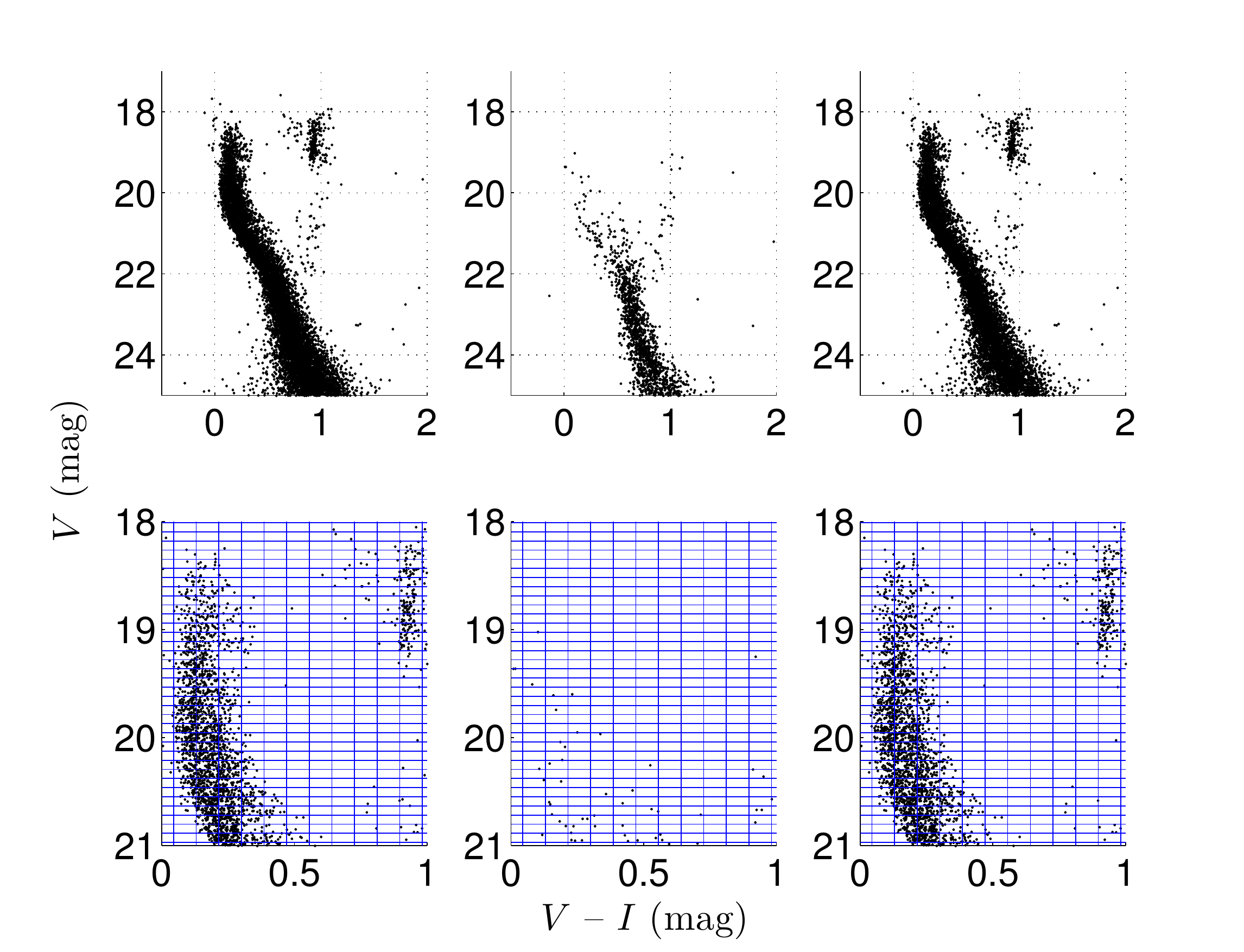}
\caption{(top left) Raw NGC 1831 CMD. (top middle) CMD of a
  representative nearby field region. (top right) Decontaminated
  cluster CMD. (bottom) Enlargements of the region containing the TO
  and RC.}
\label{fig1}
\end{figure}

We decontaminated the cluster CMDs based on an approach similar to
that applied by \cite{Kerb05}, \cite{Hu10}, and \cite{Li13}. For the
CMDs of both the cluster and the nearby field region, we generated a
common grid using 50 bins in color (spanning the range from $V-I =
-0.8$ to 3.1 mag for NGC 1831, and from $V-I = -1.3$ to 3.2 mag for
NGC 1868) and 100 bins in magnitude ($15.8 \le V \le 27.0$ mag for NGC
1831; $16.0 \le V \le 27.3$ mag for NGC 1868). We recently showed
  that varying the bin size has a trivial effect on the quality of our
  decontamination procedure \citep{Li13}. For each grid cell, we
counted the number of nearby stars per unit area. Following
  correction for the difference in area covered between the cluster
  and field regions, we then calculated the number of possible
contaminating stars in the same cell of the cluster CMD and randomly
deleted this number of stars from the relevant cell. Figure \ref{fig1}
(top panels) shows the raw NGC 1831 CMD, as well as the corresponding
CMD of the nearby field region and the decontaminated cluster CMD. The
bottom panels show zooms of the CMD region containing the TO and
RC. Only few background stars are contained in this region, so the
decontamination process is mainly important for the faint section of
the MS, which we will use to estimate the clusters' binary fractions
(see Section 3).

\subsection{Cluster Centers and Sizes}

To select a proper cluster region to examine in detail, derivation of
reliable cluster center coordinates is essential. We obtained new and
updated cluster center positions by following the same approach as
\cite{Li13}. We divided the stellar spatial distributions into 20 bins
in both right ascension ($\alpha_{\rm J2000}$) and declination
($\delta_{\rm J2000}$). In both spatial coordinates, the stellar
number densities follow Gaussian-like profiles, which allow us to
determine the coordinates corresponding to the two-dimensional maximum
density \citep[see also][]{grij13}. The resulting center coordinates
of NGC 1831 and NGC 1868 are, respectively, $\alpha_{\rm J2000} =
05^{\rm h}06^{\rm m}17.3^{\rm s}$ $(76.571707^{\circ}), \delta_{\rm
  J2000} = -64^{\circ}55'09.5''$ $(-64.919302^{\circ})$ and
$\alpha_{\rm J2000} = 05^{\rm h}14^{\rm m}35.9^{\rm s}$
$(78.649207^{\circ}), \delta_{\rm J2000} = -63^{\circ}57'12.9''$
$(-63.953577^{\circ})$. These coordinates are consistent with those
determined by \cite{Mack03}.

\begin{figure}[ht!]
\plotone{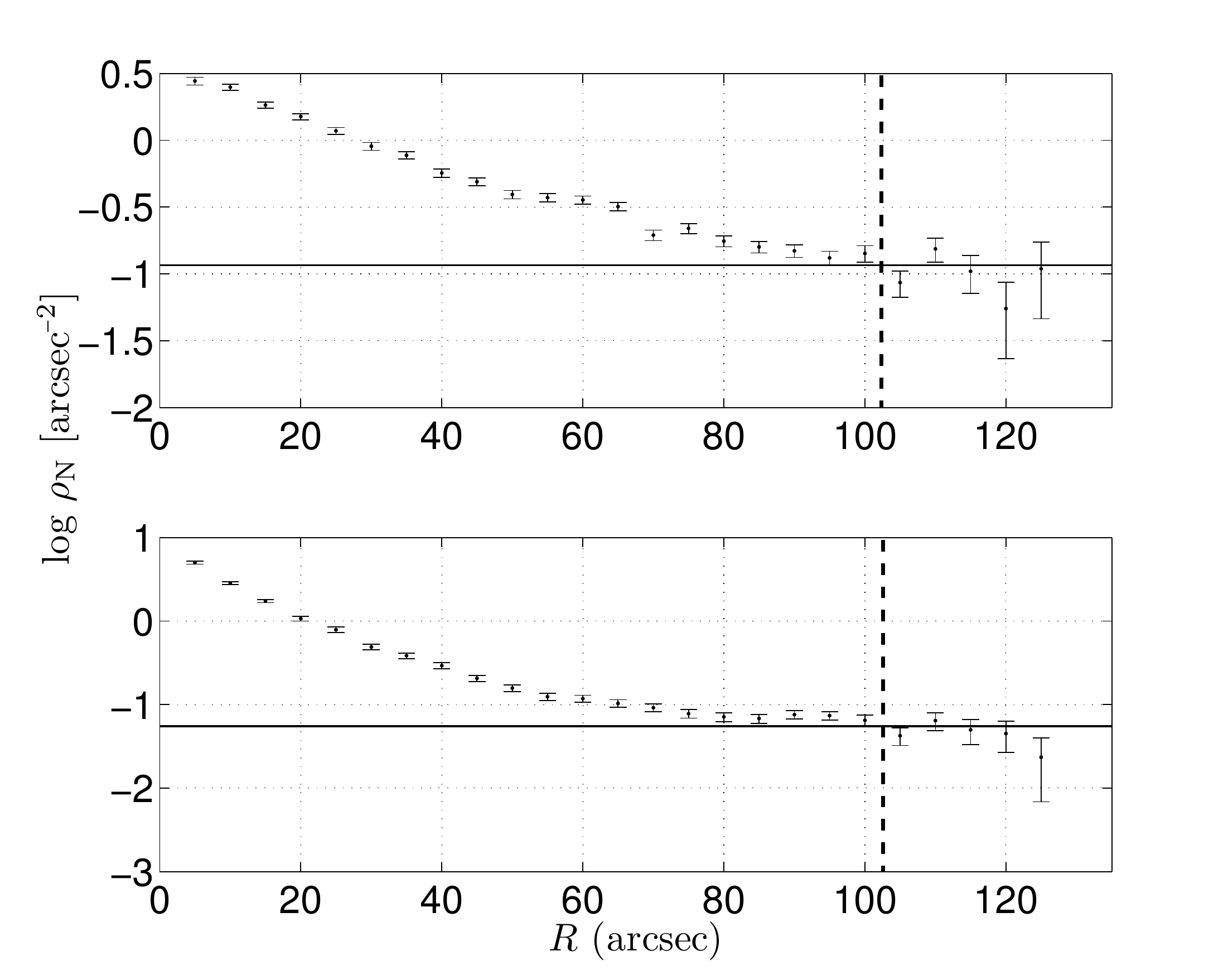}
\caption{Projected azimuthal stellar number-density profiles for (top)
  NGC 1831 and (bottom) NGC 1868, corrected for incompleteness
    and only based on stars with magnitudes brighter than the relevant
    50\% completeness limits. Black solid lines: Field-star
  levels. Black vertical dashed lines: Radii where the cluster
  densities disappear into the relevant field levels.}
\label{fig2}
\end{figure}

We next determined the (projected) radial stellar number-density
profiles for both clusters by counting the numbers of stars in
different rings, corrected for the effects of sampling
  incompleteness. We estimated the areas of the individual rings
based on Monte Carlo-type sampling, because some of the rings are
incompletely covered by the observations. For both clusters, the
projected number-density profiles monotonically decrease from the
cluster center to the periphery. The radii where the stellar number
density disappears into the average field-star level are adopted as
the clusters' sizes, $R$: $R = 102.5\pm2.5$ arcsec and $R =
103.5^{+1.5}_{-3.5}$ arcsec for NGC 1831 and NGC 1868, respectively,
 where we have only used stars with magnitudes brighter than the
  relevant 50\% completeness limits. Our sizes are systematically
larger than the equivalent values of \cite{Mack03}, who assigned their
maximum value of 76$''$ to both clusters. This is because
\cite{Mack03} limited the region used for their analysis to radii
within 75--80 arcsec, so as to avoid having to deal with very large
uncertainties. Figure \ref{fig2} displays the resulting radial
number-density profiles, as well as the levels of the field stellar
number densities we determined. Based on the estimated cluster sizes,
all NGC 1831 and NGC 1868 stars used for our analysis are constrained
to be located within 100$''$ of the cluster centers.\footnote{At the
  canonical distance to the LMC, $(m - M)_0 = 18.50$ mag, 1$''$
  corresponds to 0.26 pc.}

\section{Helium-abundance Inhomogeneities and Age Dispersion}

\begin{figure}[ht!]
\plotone{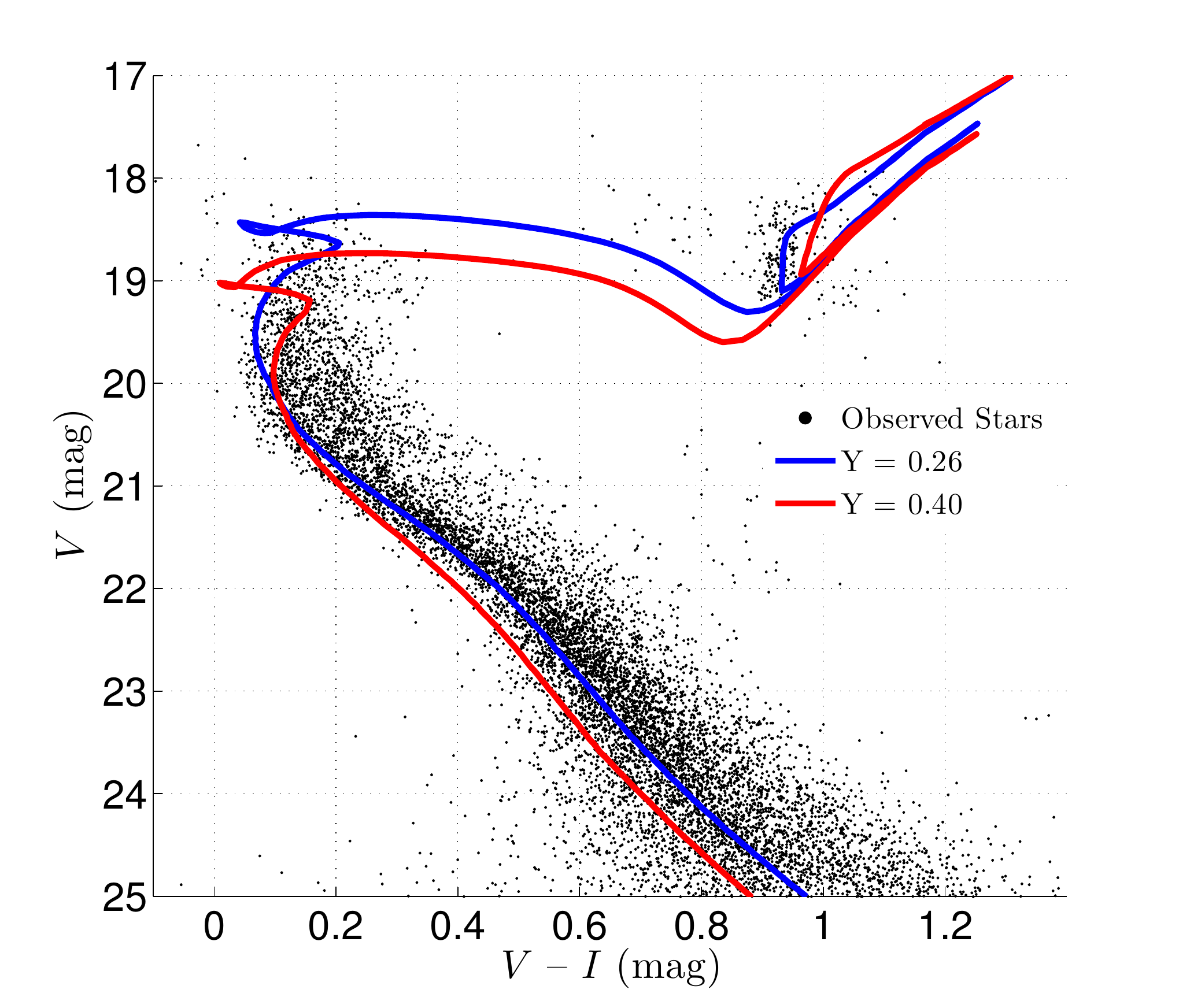}
\caption{NGC 1831 CMD, combined with the best-fitting $Y=0.26$
  isochrone (blue) and the corresponding $Y=0.40$ isochrone
  (red).}
\label{fig3}
\end{figure}

We first investigate at which level helium-abundance variations will
affect the observed extent of the MS TO. The main features of the
observed CMDs explored in this study are the following (see also
  Fig. \ref{fig3}):

\begin{enumerate}
\item A clear, well-defined MS crosses the NGC 1831 CMD from $V
  \sim 21.0, (V - I) \sim 0.35$ mag (for NGC 1868, the equivalent
  values are $V \sim 21.5, V-I \sim 0.4$ mag) to the `bottom' of the
  CMD ($V \sim 27.0$ mag, i.e., close to detection limit).
\item For approximately $V\in [21.0, 24.0]$ mag ($V\in [21.5, 24.0]$
  mag for NGC 1868), the CMD exhibits a lower-density, extended
  envelope to redder colors and (up to 0.752 mag) brighter
    magnitudes (see below), which is mainly composed of unresolved
  binary systems \citep[for more details,
    see][]{Rube97,Elso98,Hu10,grij13,Li13}. The properties of this
  `binary envelope' allow us to estimate reliable binary fractions for
  our two sample clusters, which we use as an input parameter in our
  analysis.
\item For $V\leq21.0$ (21.5) mag, the MS broadens and the distinct
  boundary between the MS and its binary envelope disappears. The MS
  turns into an extended MS TO region. The color dispersion of the MS
  TO region reaches $\geq$ 0.2 mag, which cannot be simply owing to
  photometric uncertainties.
\item Both NGC 1831 and NGC 1868 show a RC which is extended in
  magnitude but compact in color: $V, (V - I) \sim$ [18.2 to 19.2,
    $\sim 0.9$] mag for NGC 1831 and $V, (V - I) \sim$ [18.7 to 19.7,
    $\sim 0.9$] mag for NGC 1868.
\end{enumerate}

To quantitatively estimate the effects of helium-abundance variations,
we first use an isochrone characterized by the standard helium
abundance, $Y=0.26$, to obtain an adequate, visual fit to the
observed CMD. We only take NGC 1831 as an example here, because we
will show that assuming a dispersion in helium abundance is
inconsistent with the observed, narrow MS; in fact, for NGC 1868 we
find the same inconsistency. For NGC 1831, we adopt an age of $\log(t
\mbox{ yr}^{-1}) = 8.80$ here (we will vary the cluster age in the
remainder of our analysis), a metallicity of $Z = 0.012$ (where
$Z_\odot = 0.019$), which is identical to that derived by
\cite{Kerb05}, and a total extinction of $E(B-V) = 0.03$ mag. Our
adopted age is within the range quoted by \cite{Kerb06}, $8.70 \le
\log(t \mbox{ yr}^{-1}) \le 8.95$, while our extinction estimate is
consistent with that of \cite{Javi05}, although it is slightly higher
than that derived by \cite{Liu09}, $E(B-V) = 0.00$ mag. This latter
apparent discrepancy is driven by our choice of fitting the MS
ridgeline rather than the bottom of the MS envelope, as done by
\cite{Liu09}; see below. The isochrone characterized by these
parameters matches the ridgeline of the faint section of the NGC 1831
(NGC 1868) MS, $V\leq21.0$ (21.5) mag, as well as the blue boundary of
the extended MS TO and the compact RC: see Fig. \ref{fig3} (blue
isochrone). We use the isochrone database of \cite{Bert08,Bert09},
obtained from the YZVAR interactive web
service,\footnote{http://stev.oapd.inaf.it/YZVAR/cgi-bin/form}
specifically to generate isochrones for different helium
abundances. However, this database cannot be used to generate
isochrones in the {\sl HST}/WFPC2 photometric system. Instead, we
generate isochrones in the {\sl HST}/Advanced Camera for Surveys
(ACS)-High-Resolution Camera (HRC) photometric system. Following
\cite{Siri05}\footnote{See their Eq. (12), tables 11 and 21.} we
convert the ACS-HRC isochrones to the {\sl HST}/WFPC2 system.

However, even if we increase the helium abundance to $Y=0.40$, the
corresponding isochrone still hardly contributes to the generation of
an extended TO (in combination with the $Y = 0.26$ isochrone). More
importantly, isochrones characterized by $Y=0.40$ define a clearly
bluer MS than the equivalent $Y=0.26$ isochrones, and a redder RC, but
the observed narrow MS and compact RC invalidate any suggestion of a
significant spread or significant inhomogeneities in helium abundance.

Next, we explore whether the age-dispersion model -- for the same
$Y,Z$ chemical abundance -- can satisfactorily match our observational
data. We use the \cite{Bres12} isochrones to match the decontaminated
CMDs, because these isochrones are up to date and include the {\sl
  HST}/WFPC2 filter set \citep[as opposed to the][isochrones used
  above]{Bert08,Bert09}. First, we focus on the extended MS TO
region. For most intermediate-age LMC clusters, the age-dispersion
model has been used to show that their extended MS TOs can be
described accurately by two isochrones with an age difference of $\sim
300$ Myr. We determine the isochrones that best match the blue
(younger) and red (older) outer boundaries of the (presumably single)
MS stars. For NGC 1831, both isochrones cross at $[V,(V-I)] \sim
[21.0, 0.35]$ mag ($[V,V-I] \sim [21.5, 0.40]$ mag for NGC 1868). The
best-fitting age dispersion for NGC 1831 covers the range from $\log(t
\mbox{ yr}^{-1}) = 8.74$ to 8.92, i.e., roughly 280 Myr. For NGC 1868,
the age range covered runs from $\log(t \mbox{ yr}^{-1}) = 8.93$ to
9.07, again indicating a potential age dispersion of roughly 320
Myr. Thus, the age dispersion implied by our isochrone fits is
consistent with the average age dispersion reported previously for
intermediate-age LMC clusters ($\sim 300$ Myr).

\begin{figure}[ht!]
\plotone{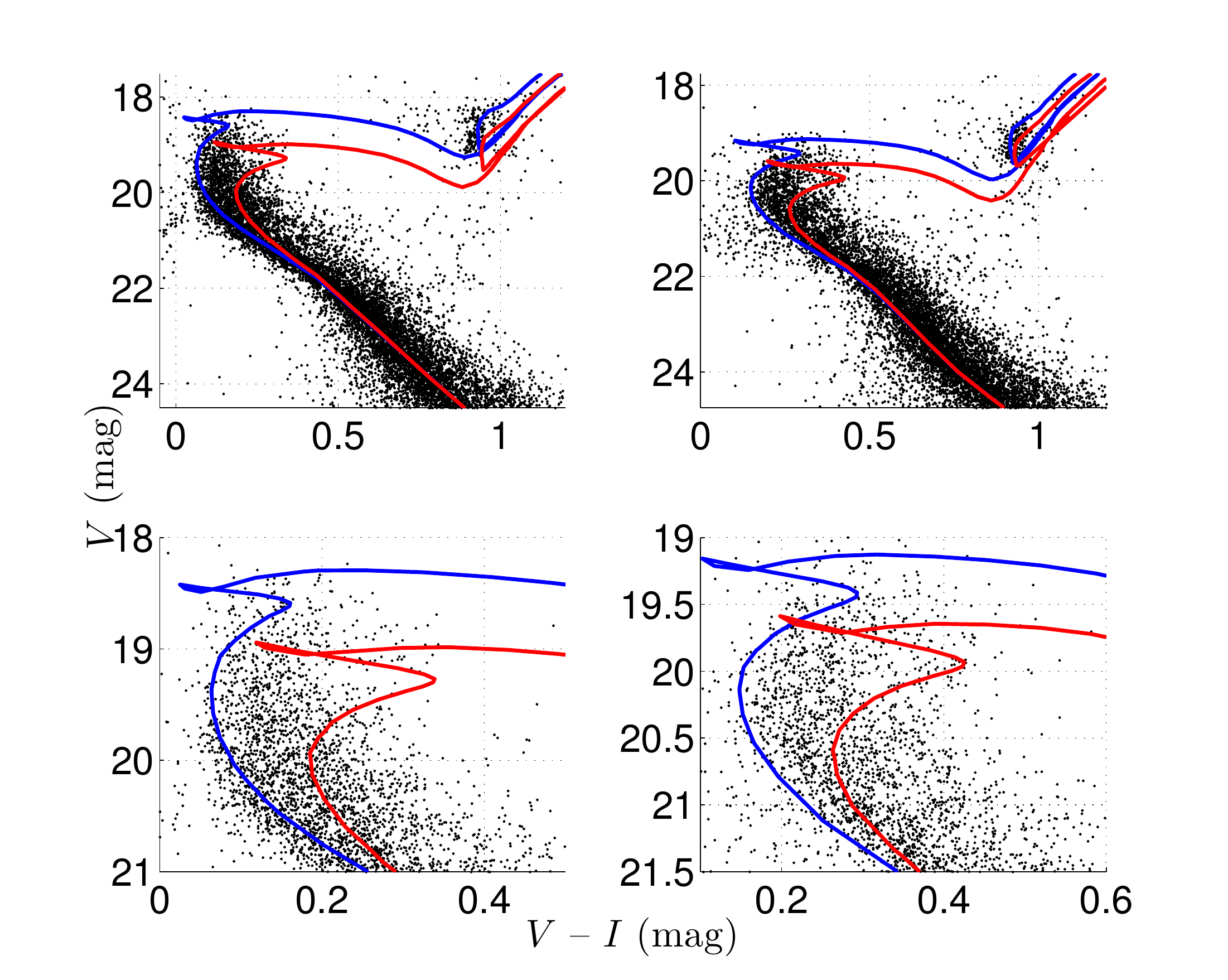}
\caption{(top left) NGC 1831 CMD and best-fitting isochrones for
  $\log(t \mbox{ yr}^{-1}) = 8.74$ (blue) and 8.92 (red). (top right)
  NGC 1868 CMD and best-fitting isochrones for $\log(t \mbox{
    yr}^{-1}) = 8.93$ (blue) and 9.07 (red). (bottom) Enlargements of
  the regions covering the extended MS TOs of (left) NGC 1831 and
  (right) NGC 1868.}
\label{fig4}
\end{figure}

\begin{figure}[ht!]
\plotone{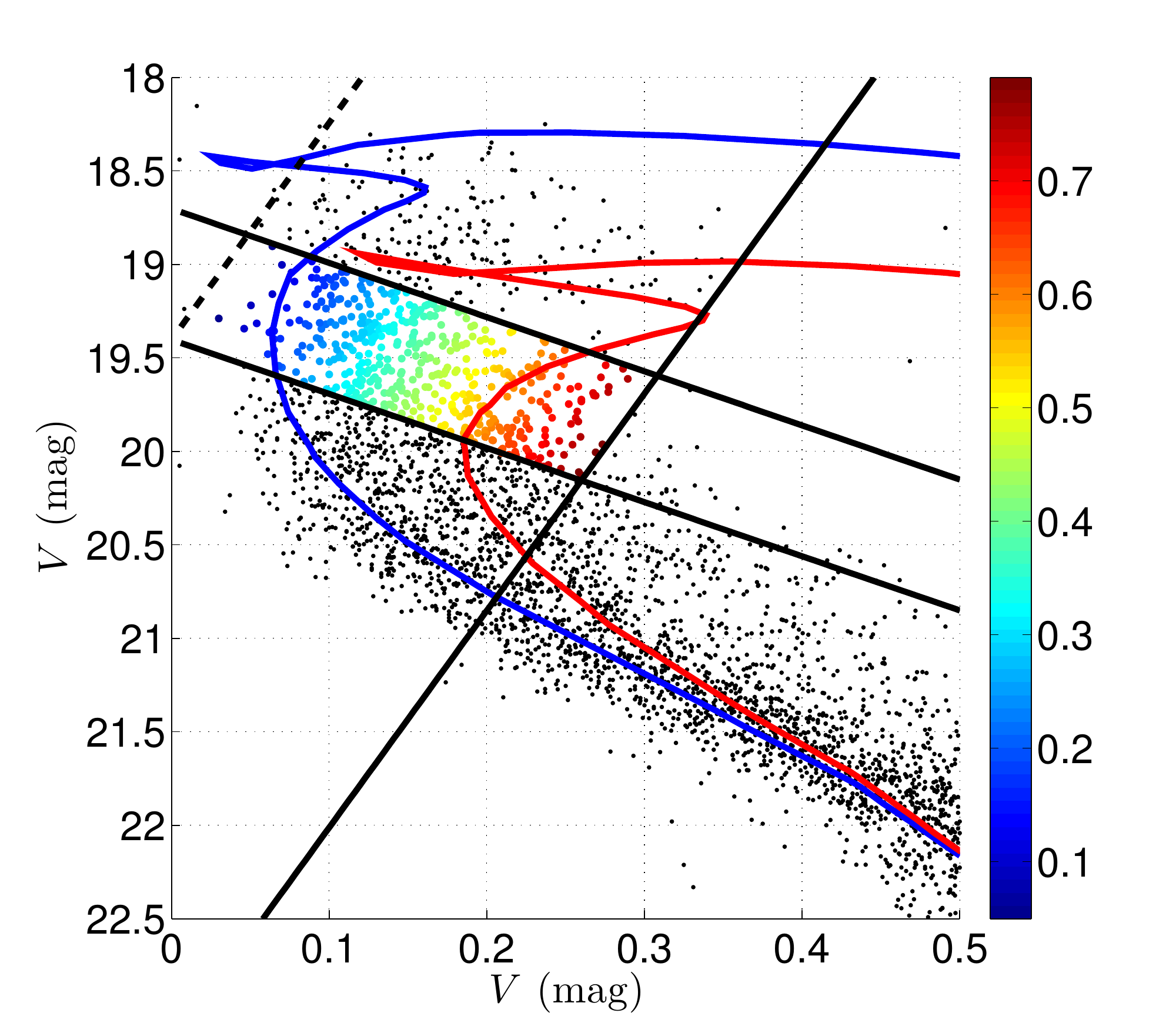}
\caption{Illustration of the method used to determine the pseudo-color
  distribution of NGC 1831 MS TO stars. The region of interest is
  delineated by three black solid lines and a fourth, dashed line. We
  determined the minimum distance of the stars contained in this
  region to its left-hand boundary (black dashed line), which we
  define as the `pseudo-color.' The blue and red lines represent the
  isochrones adopted as young and old limits. The color scale
  indicates the relevant magnitudes along the pseudo-color axis.}
\label{fig5}
\end{figure}

\begin{figure}[ht!]
\plotone{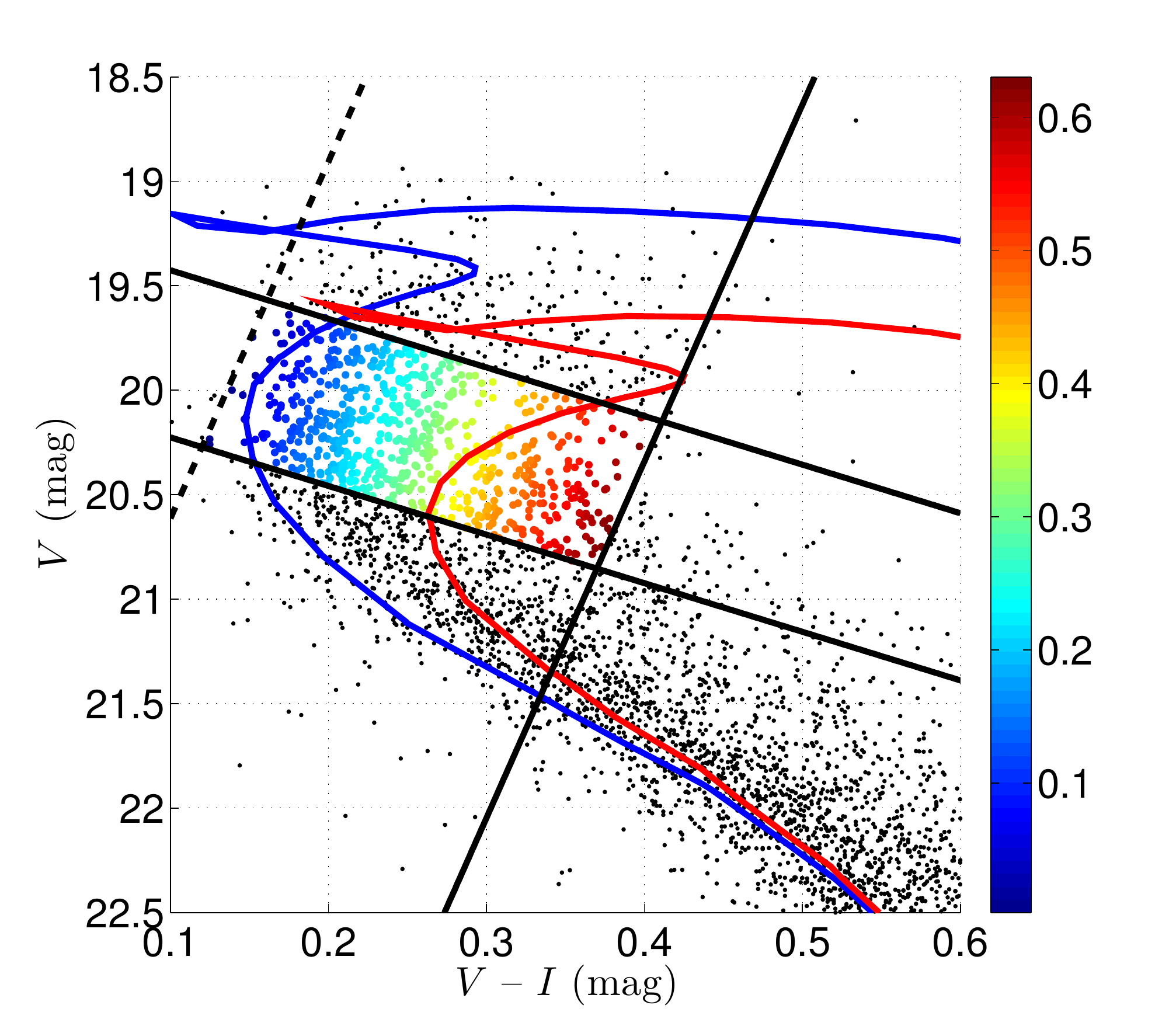}
\caption{As Fig. \ref{fig5}, but for NGC 1868.}
\label{fig6}
\end{figure}

\cite{Kerb06} determined the youngest age boundaries for NGC 1831 and
NGC 1868 at $\log(t \mbox{ yr}^{-1}) = 8.70$ and 8.95, respectively,
which matches our lower boundaries very well. For NGC 1831, we adopted
a metallicity of $Z=0.012$, an extinction of $E(B-V) = 0.03$ mag, and
a distance modulus of $(m-M)_0 = 18.40$ mag. Adoption of these
parameters results in an isochrone set that closely describes the
cluster's MS ridgeline (see Figs \ref{fig5} and \ref{fig6}). For NGC
1868, the equivalent values are $Z=0.008, E(B-V)=0.04$ mag, and
$(m-M)_0=18.48$ mag. The metallicities we adopted are identical to
those assumed by \cite{Kerb06}. However, compared with their adopted
distance modulus -- $(m-M)_0 = 18.70$ mag -- our results yield
distance moduli that are closer to the LMC's canonical value of 18.50
mag. We also compared our adopted $(m-M)_0$ with those derived by
\cite{Liu09}, who found best-fitting values of 18.58 and 18.55 mag for
NGC 1831 and NGC 1868, respectively. By checking their isochrone fits
in detail (their fig. 1), we find that \cite{Liu09} adopted isochrone
fits that describe the bottom of the MS, while we adopted isochrones
that match the MS ridgeline instead. Our adopted $E(B-V)$ values are
also slightly higher than those determined by \cite{Liu09}---0.00 and
0.02 mag for NGC 1831 and NGC 1868, respectively---which is caused by
the same difference in fitting approach. We have summarized all
cluster parameters in Table 1. Figure \ref{fig4} shows the cluster
CMDs as well as the isochrones adopted. We also provide zooms of the
region covering the extended MS TO.

If we simplistically assume that the extended MS TOs of our sample
clusters are caused by age dispersions, the stellar distribution in
the MS TO regions should reflect the stellar populations' age
distributions. For example, double MS TOs may indicate the occurrence
of two distinct starburst events, while extended MS TOs would be more
appropriately described by a scenario based on continuous star
formation. To constrain the clusters' age distributions, we explored
the `pseudo-color' distributions of the stars found near the MS TOs,
as follows:

\begin{enumerate}
\item We defined a region containing a subset of MS TO stars that
  are not contaminated by subgiant stars. The adopted young and old
  isochrones define the magnitudes and colors of their respective MS
  TO points, which we connected using a straight line. Using this line
  as one boundary to the region of interest, we slide the young MS TO
  to a locus that is close to the blue boundary defined by the MS TO
  stars. From this point, we generate a roughly tangential final
  boundary coincident with the blue envelope of the MS TO stars: see
  Figs \ref{fig5} and \ref{fig6}. We explored a number of options
    to define the most appropriate region for our analysis of the
    presence of a possible age gradient, ranging from thin, $\la 0.5$
    mag-wide `shells' along the pseudo-color vector similar to those
    adopted by \cite{Goud11} to the full extent of the regions
    indicated in Figs \ref{fig5} and \ref{fig6}. The resulting
    pseudo-color (and the associated age) distributions proved to be
    similar no matter where we set the region's upper limit; we chose
    as upper limit for the remainder of our analysis the brightest
    magnitudes just below the subgiant branch (where we have taken
    into account the relevant photometric uncertainties) determined by
    the oldest isochrone so as to achieve the optimum statistical
    significance.
\item For each star in this region, we calculate their minimum
  `distance' to the region's left-hand boundary, which is tangent to
  the blue edge of the extended MS TO region (see the black dashed
  lines in Figs \ref{fig5} and \ref{fig6}). This distance represents
  the pseudo-color for each MS TO star.
\item Finally, we obtained the distribution of the stars in this
  region along the pseudo-color direction, normalized to the total
  number of stars in the region.
\end{enumerate}

\begin{figure}[ht!]
\plotone{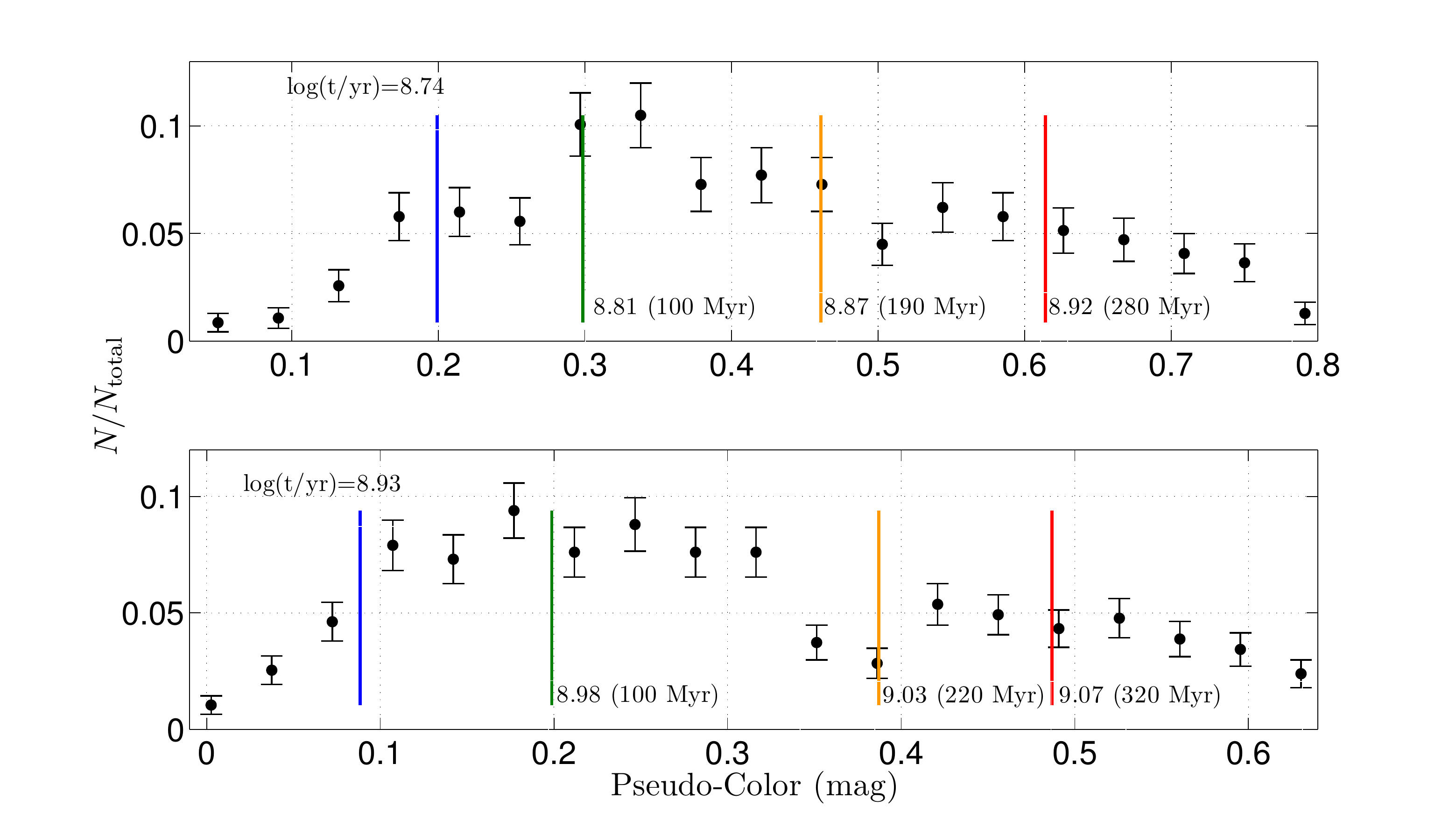}
\caption{Normalized pseudo-color distributions of the MS TO stars in
  our sample clusters. (top) NGC 1831. (bottom) NGC 1868. The
    vertical, colored lines indicate the ridgelines of the isochrones
    representative of the ages indicated in the figure.}
\label{fig7}
\end{figure}

\begin{figure}[ht!]
\plotone{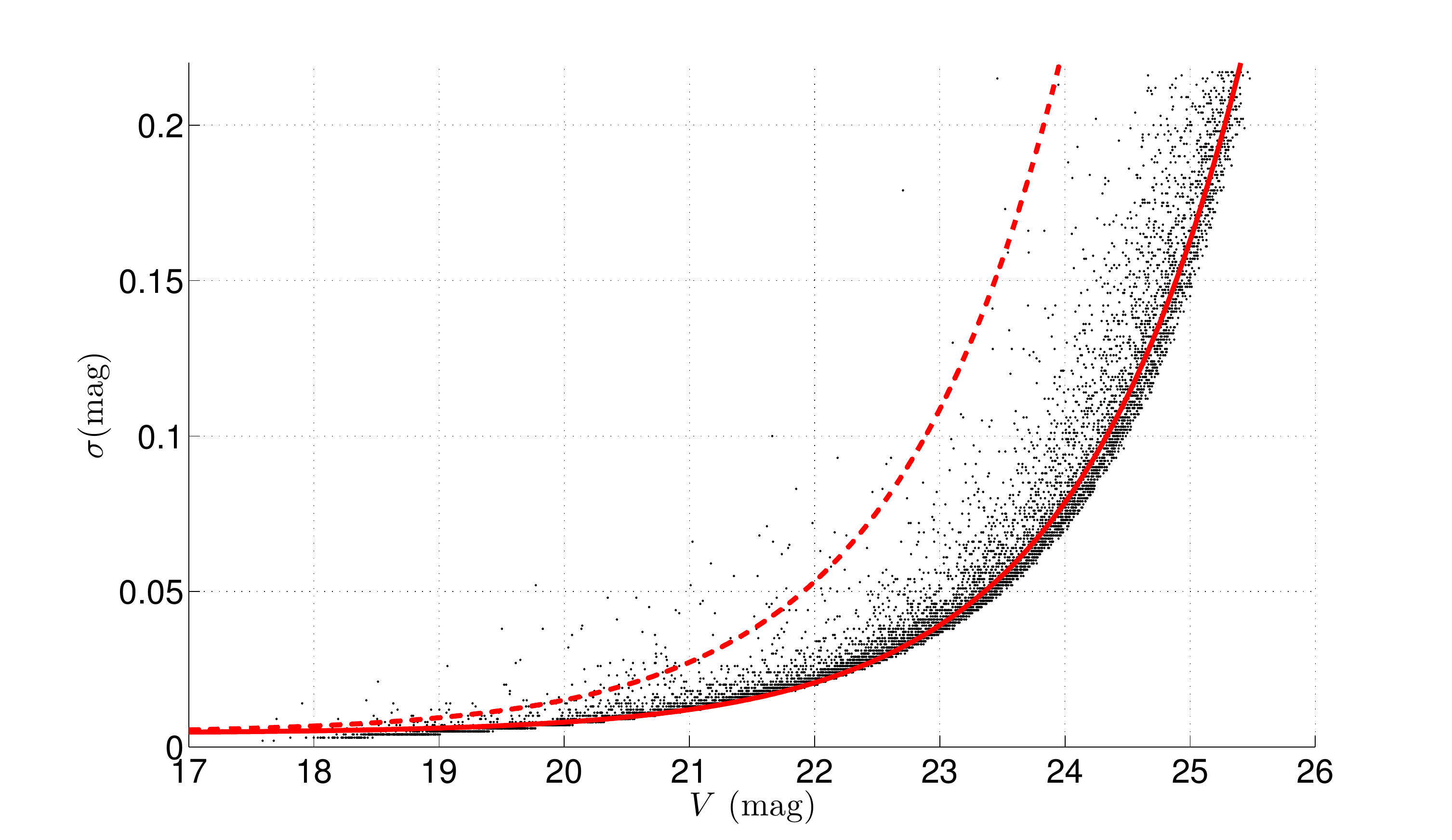}
\caption{Photometric uncertainties as a function of stellar
  magnitude. The red solid line indicates the best-fitting functional
  form of the correlation, while the red dashed line indicates the
  relevant $3\sigma$ level, for NGC 1868.}
\label{fig8}
\end{figure}

The resulting pseudo-color distributions are shown in
Fig. \ref{fig7}. For both clusters, the MS TO stars' pseudo-color
spread can reach 0.6--0.8 mag, which again cannot be simply due to
photometric uncertainties (see Fig \ref{fig8}). Both clusters show
clear peaks in their pseudo-color distributions, at $\sim 0.30$ mag
and 0.18 mag for NGC 1831 and NGC 1868, respectively. The pseudo-color
distributions also exhibit broad red wings. Part of this broadening
may be owing to the presence of unresolved binaries.  In the
  selected regions in Figs \ref{fig5} and \ref{fig6}, different
  isochrones cover different pseudo-colors. One can define a range of
  ages in order to explore the possible presence of an age gradient
  across the regions. In Fig. \ref{fig7}, we show the results for four
  appropriate, distinct ages (see labels), which divide the entire
  allowed age dispersion into four roughly equal ranges. Note that the
  isochrones in the selected region are not exactly straight lines, so
  that these distinct age steps are only indicative of the positions
  of the theoretical ridgelines represented by the isochrones
  labeled.

At the distance to the LMC, most binary systems in compact clusters
are unresolved in observations with current instruments. The
magnitudes of these unresolved binary systems will thus be composed of
the co-added fluxes of the individual components:
\begin{equation}
  m_{\rm b}=-2.5 \log(10^{-0.4m_1}+10^{-0.4m_2}), 
\end{equation}
where $m_1$ and $m_2$ are the component magnitudes. Binaries which are
composed of two identical components will exhibit the same color but
be brighter by 0.752 mag than the individual stars
themselves. Unresolved binaries composed of individual member stars
that are not identical will exhibit shifts in both color and
magnitude. The resulting color will be redder than that of the primary
star but bluer than that of the secondary component. 

\begin{figure}[ht!]
\plotone{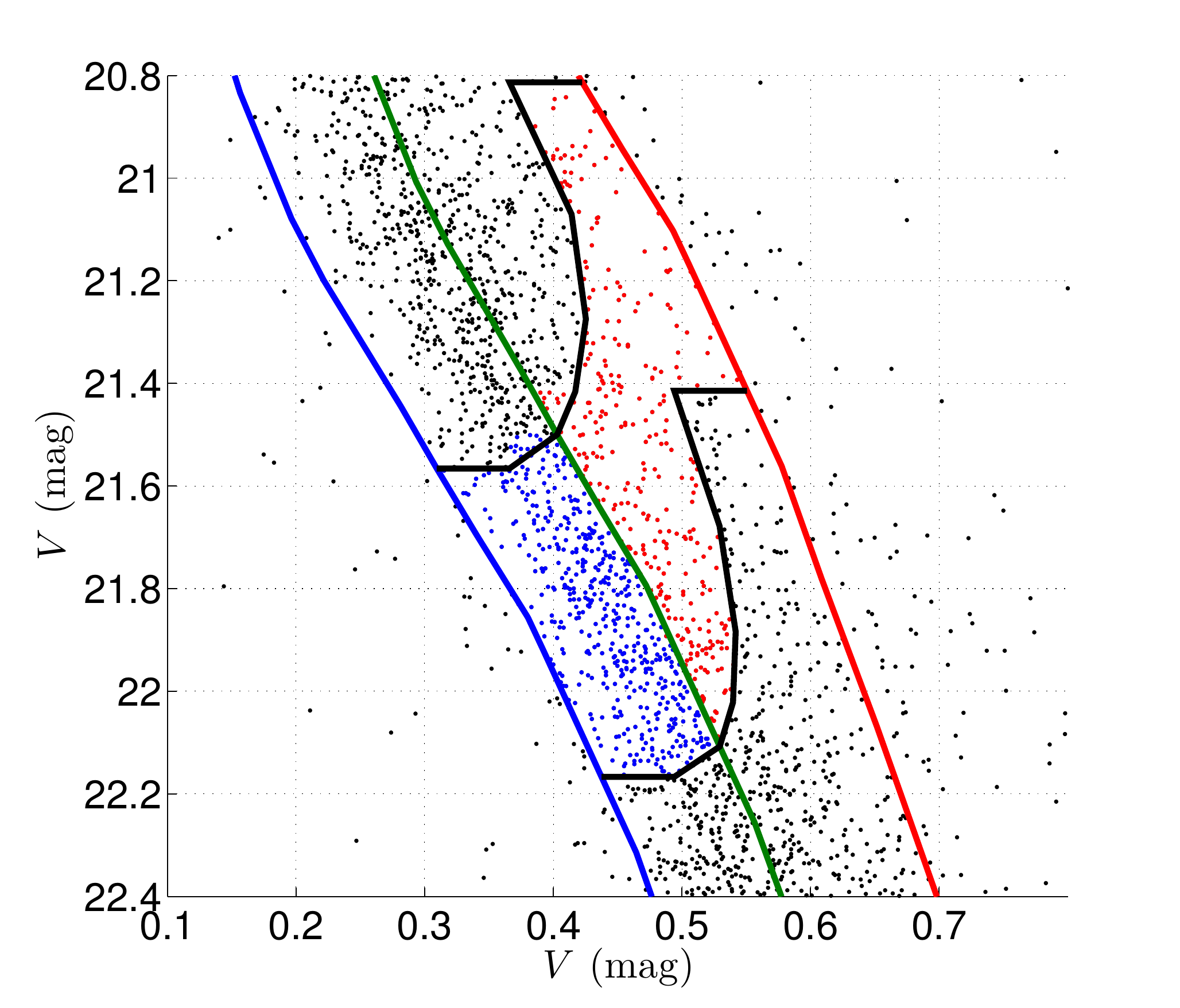}
\caption{Adopted binary- and single-star regions along the MS for NGC
  1868. See text for a detailed description of our approach.}
\label{fig9}
\end{figure}

In this paper, the approach we used to determine the clusters' binary
fractions is similar to that employed by \cite{Milo11}. In
Fig. \ref{fig9}, we take NGC 1868 as an example. For unresolved
binaries characterized by different mass ratios -- $q = m_2/m_1$ --
the resulting photometry will be biased to brighter magnitudes and
redder colors as $q$ increases from 0 to 1, thus resulting in a
statistical broadening of the MS towards the brighter and redder side
\citep[see also][]{Hu10,Milo11,grij13,Li13}. For the brightest section
of the MS, the possible presence of an age dispersion will blur any
features due to binarity, so the bright magnitude range cannot be used
to determine the clusters' binary fractions. The faintest section of
the observed MS is not an ideal range to explore the quantitative
contributions of binaries either, because the prevailing large(r)
photometric uncertainties cause contamination due to a mixture of
single stars and binaries.

Thus, we can only use MS stars of intermediate magnitudes to determine
the binary fractions. For NGC 1868, we only select primary stars
between $V= 21.6$ mag and 22.2 mag for further analysis of their
binary populations. For NGC 1831, the equivalent magnitudes range from
$V = 21.4$ mag to 21.9 mag. {\sc HSTphot} also yields the photometric
uncertainties for all stars. We use an exponential function of the
form $\sigma({\rm m}) = \exp(a\times{\rm m}+b)+c$ to fit the
photometric uncertainties as a function of stellar magnitude,
following \cite{Hu10} and \cite{Li13}. The best fit is shown in
Fig. \ref{fig8}. Using the equation thus obtained, we adopt as the
blue boundary of the MS range for further analysis the locus of the
best-fitting isochrone, adjusted by the prevailing $-3\sigma$
photometric uncertainties of the stellar distribution in the relevant
magnitude range (see Fig. \ref{fig9}). We show `binary isochrones' for
different mass ratios, from $q = 0$ to unity (corresponding to the
single-star isochrone shifted by $-0.752$ mag). The locus of the $q =
1$ binary isochrone $+3\sigma$ uncertainty is the red boundary of the
region of interest for this analysis (see Fig. \ref{fig9}). Both
boundaries, as well as the black solid lines connecting the different
values of $q$, define the region we used to count binaries and single
stars.

Because of the prevailing photometric uncertainties, in \cite{grij13}
and \cite{Li13}, we adopted $q = 0.55$ as the critical locus to
distinguish between single stars and binaries. In this paper, we
increase this threshold to $q=0.6$ (see the dark green line in
Fig. \ref{fig9}), because the MS magnitude range of interest is
fainter than that used by both \cite{grij13} and \cite{Li13}. Finally,
adopting the constraints given by the black solid lines, we adopted
stars located within the region from $-3\sigma$ of the best-fitting
isochrone to the $q = 0.6$ binary isochrone as single stars (shown as
the blue dots in Fig. \ref{fig9}), while stars that lie beyond the $q
= 0.6$ binary isochrone but with colors bluer than the $q = 1.0$
isochrone $+3\sigma$ were assumed to be binaries (the red dots in
Fig. \ref{fig9}). The binary fraction resulting from this procedure,
$f_{\rm b} = \frac{N_{\rm b}}{(N_{\rm b}+N_{\rm s})}$, is 29.2\% for
NGC 1831 and 33.5\% for NGC 1868. These values are similar to
  those of \cite{Soll10}, who analyzed five Galactic open clusters
  using a similar approach as adopted in this paper, and found cluster
  core binary fractions between 11.9\% and 34.1\%. These fractions
  are, however, significantly higher than those of \cite{Milo12}, who
  derived the (core) binary fractions of 59 Galactic globular
  clusters. Their maximum core binary fraction only reaches 17\%. This
  is because they mainly investigated old globular clusters, so that
  primordial binaries will likely have been disrupted. Indeed, in
  \cite[][their fig. 8]{Li13} we found binary fractions (for a similar
  magnitude range) of $\sim35$\% and 30\% for the young massive
  clusters NGC 1805 and NGC 1818, respectively.

Note that these binary fractions only apply to $q\geq0.6$
binaries. Occasional blending of stars along the line of sight will
introduce a small bias. We assume a flat mass-ratio distribution and
use a Monte Carlo method to correct for this blending bias \citep[for
  more details, see][]{Hu10,grij13,Li13}. This leads to a final
estimate of the total binary fraction, for all mass ratios, of 67.2\%
for NGC 1831 and 76.9\% for NGC 1868.

In the following simulation, we simply assume binary fractions of 70\%
for NGC 1831 and 75\% for NGC 1868. Even though the estimated binary
fractions are strictly only valid for the magnitude range from $V=
21.4$ (21.6) mag to $V = 21.9$ (22.2) mag for NGC 1831 (NGC 1868), we
make the assumption that these values represent the global binary
fractions, i.e., the binary fractions characterizing the clusters as a
whole (see below). Adoption of the observational level of photometric
noise will cause faint binaries with low $q (q \leq 0.6)$ to almost
all mix with and become indistinguishable from single stars. For the
TO region, the observed age dispersion blurs the expected broadening
owing to the presence of binaries. We checked that the simulated
luminosity function is similar to its observed counterpart, which
hence implies that our adoption of the measured binary fractions as
representative of the clusters as a whole is appropriate.

\begin{figure}[ht!]
\plotone{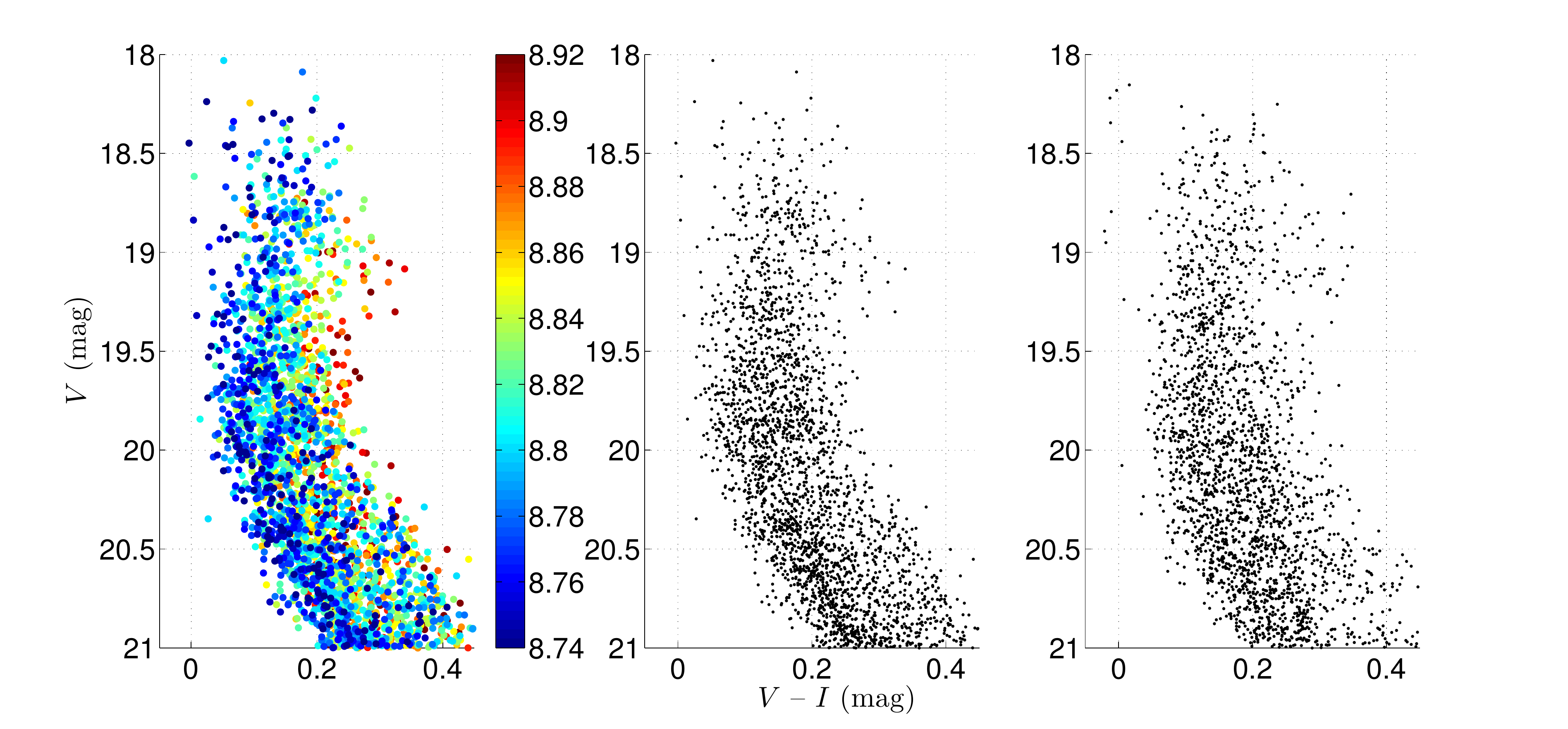}
\caption{Illustration of how we synthesized the simulated NGC
  1831-like cluster. (left) Colors are used to indicate the ages of
  the simulated stars, in units of $\log(t \mbox{ yr}^{-1})$. At faint
  magnitudes, all stars are fully mixed. A clear feature indicating
  binarity is seen as the broadening of the MS; a bifurcation appears
  close to the MS TO region (which hence contributes to an extended MS
  TO). (middle) Synthesized CMD of the simulated cluster in the left
  panel without age indications. (right) Observed CMD of NGC 1831.}
\label{fig10}
\end{figure}

\begin{figure}[ht!]
\plotone{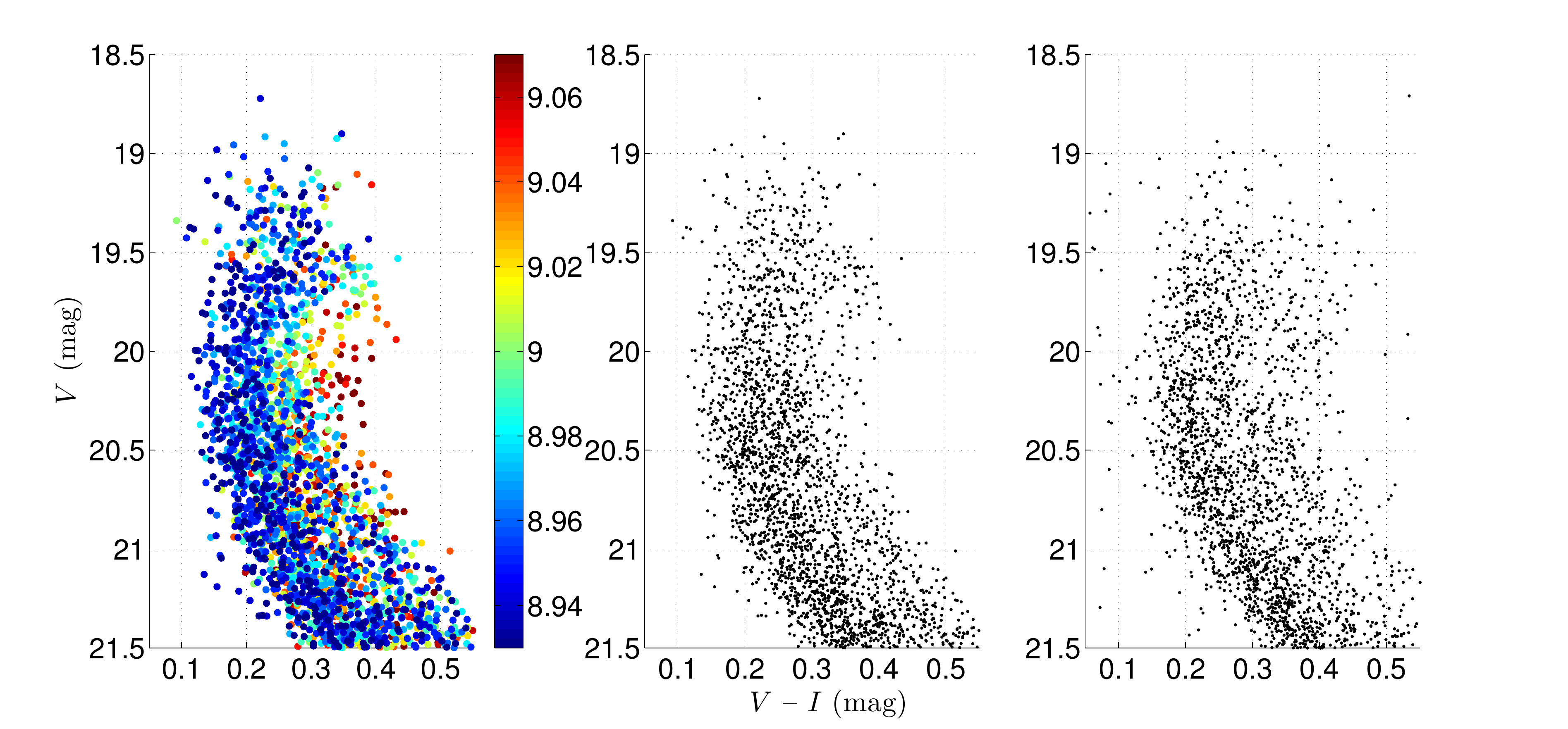}
\caption{As Fig. \ref{fig10}, but for NGC 1868.}
\label{fig11}
\end{figure}

We next performed simulations of clusters characterized by different
age dispersions and binary fractions. For both NGC 1831 and NGC 1868,
we simulated a series of clusters with different ages but
characterized by the same binary fractions. Our simulated clusters
have similar numbers of member stars, binary fractions, and extended
TOs as the observed clusters. We subsequently employed the same
approach as illustrated in Figs \ref{fig5} and \ref{fig6} to find the
pseudo-color distribution of the simulated MS TO stars. By adjusting
the stellar age distributions of the simulated clusters, we determined
the distributions that most closely reproduce the observed
pseudo-color distributions. Figures \ref{fig10} and \ref{fig11} show
the best-fitting simulated cluster CMDs for NGC 1831 and NGC 1868,
respectively. The relevant adopted input parameters are summarized in
Table 1.

\begin{table}
 \centering
 \begin{minipage}{110mm}
  \caption{Basic input parameters for our cluster
    simulations. (Numbers in brackets refer to the observations.)}
  \begin{tabular}{@{}lcc@{}}
  \hline
  Parameter                                       &  NGC 1831   & NGC 1868 \\
 \hline
 Youngest age boundary, $\log(t \mbox{ yr}^{-1})$ & 8.74        & 8.93 \\
 Oldest age boundary, $\log(t \mbox{ yr}^{-1})$   & 8.92        & 9.07 \\
 Age steps, $\Delta \log(t \mbox{ yr}^{-1})$      & 0.01        & 0.01 \\
 Number of stars ($V\leq 21.5$ mag)               & 3647 (3413) & 2487 (2273) \\
 Adopted binary fraction (\%)                     & 70          & 75 \\
 Extinction, $E(B-V)$ (mag)                       & 0.03        & 0.04 \\
 Metallicity, $Z$ ($Z_{\odot}=0.019$)             & 0.012       & 0.008 \\
\hline
\end{tabular}
\end{minipage}
\end{table}

\begin{figure}[ht!]
\plotone{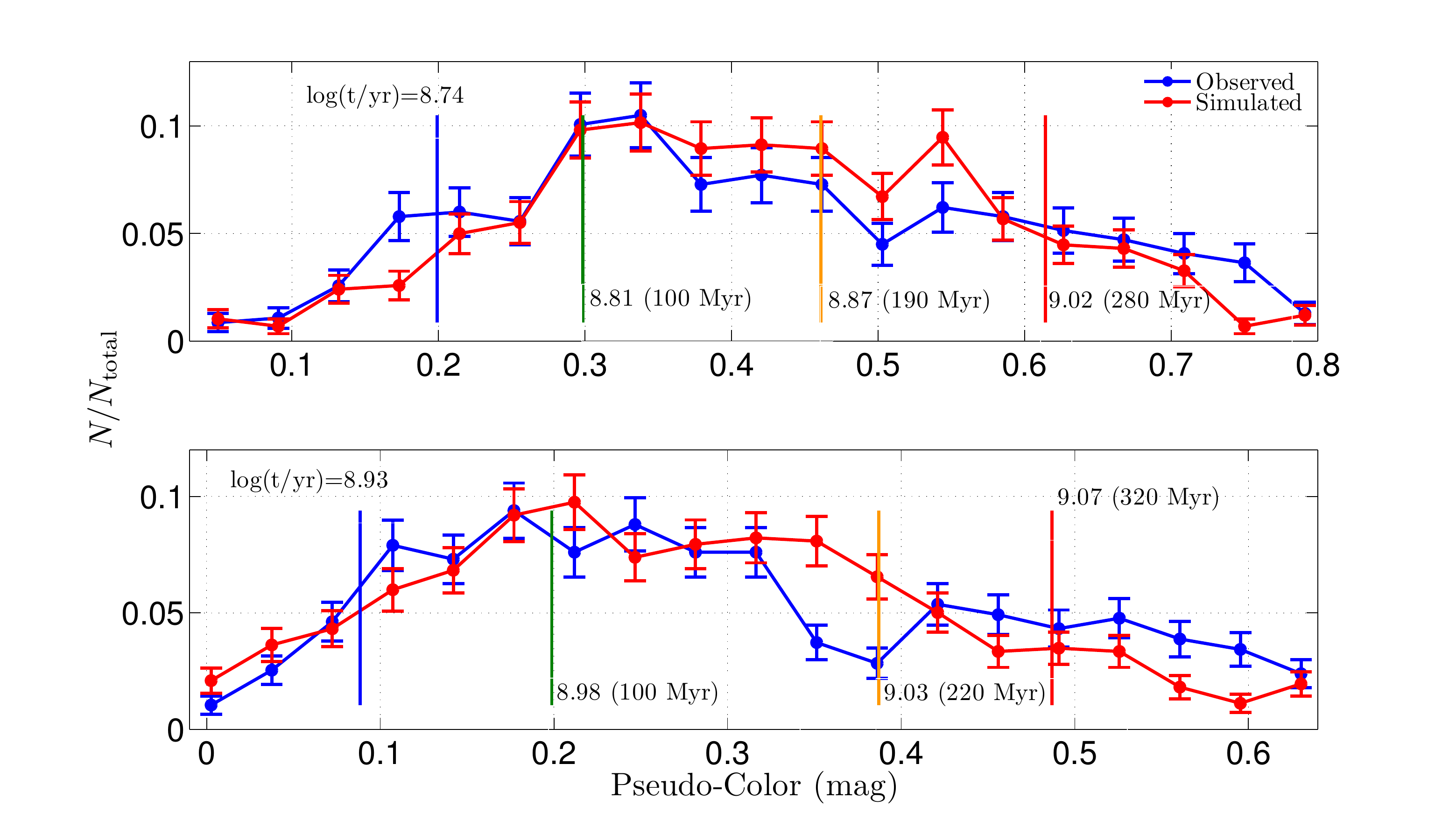}
\caption{(top) As Fig. \ref{fig7}, showing a comparison of the
  simulated (red) and observed (blue) NGC 1831 pseudo-color
  distributions. (bottom) As the top panel, but for NGC 1868.}
\label{fig12}
\end{figure}

\begin{figure}[ht!]
\plotone{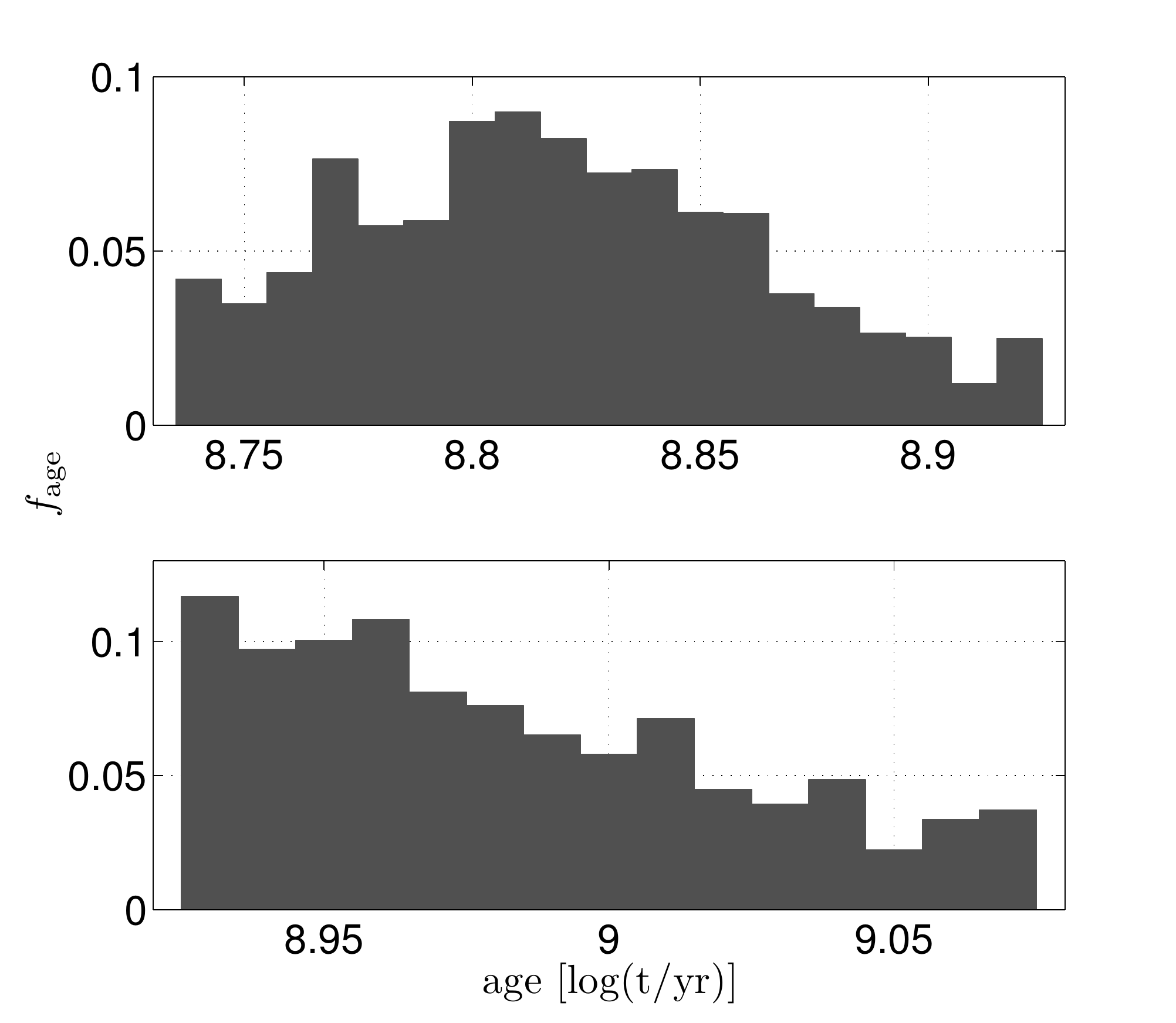}
\caption{Best-fitting age distributions for (top) NGC 1831 and
  (bottom) NGC 1868.}
\label{fig13}
\end{figure}

The resulting pseudo-color distributions for the simulated clusters
equivalent to NGC 1831 and NGC 1868, as well as a comparison with the
observations (Fig. \ref{fig7}), are shown in Fig. \ref{fig12}. The
corresponding best-fitting age distributions for our two clusters are
displayed in Fig. \ref{fig13}. For NGC 1831, the best-fitting age
distribution peaks at $\log(t \mbox{ yr}^{-1})=8.80$, with gradual
decreases towards either side. The age distribution pertaining to NGC
1868 does not exhibit a clear peak, but instead shows a gradual
decrease from $\log(t \mbox{ yr}^{-1})= 9.07$ to 8.93. We quantify the
similarities of the observed and simulated pseudo-color distributions
using a Kolmogorov--Smirnov (K--S) test.\footnote{The K--S test
  returns two values, $H$ and $P$. $H$ can only attain the two
  concrete values 0 and 1, which depend on the null hypothesis that
  the two underlying samples are independent. If $H$ is returned as 0,
  the null hypothesis is rejected, which means that the two samples
  are drawn from the same distribution. $P$ represents the $p$ value,
  which indicates the probability that the two samples are drawn from
  the same underlying population.} Our K--S evaluation shows that the
simulated pseudo-color distributions are indeed drawn from the same
underlying distributions ($H=0$), for both clusters. For both NGC 1831
and NGC 1868, the K--S test returns {$P= 74$\%}. Figure \ref{fig12}
shows that the simulated pseudo-color distribution pertaining to the
NGC 1831 simulation (top panel) almost perfectly fits that of the
observed sample. For NGC 1868, a small offset is seen on the red
side. Nevertheless, across the full range, the K--S test implies that
the distributions are indiscernible for both clusters.

\begin{figure}[ht!]
\plotone{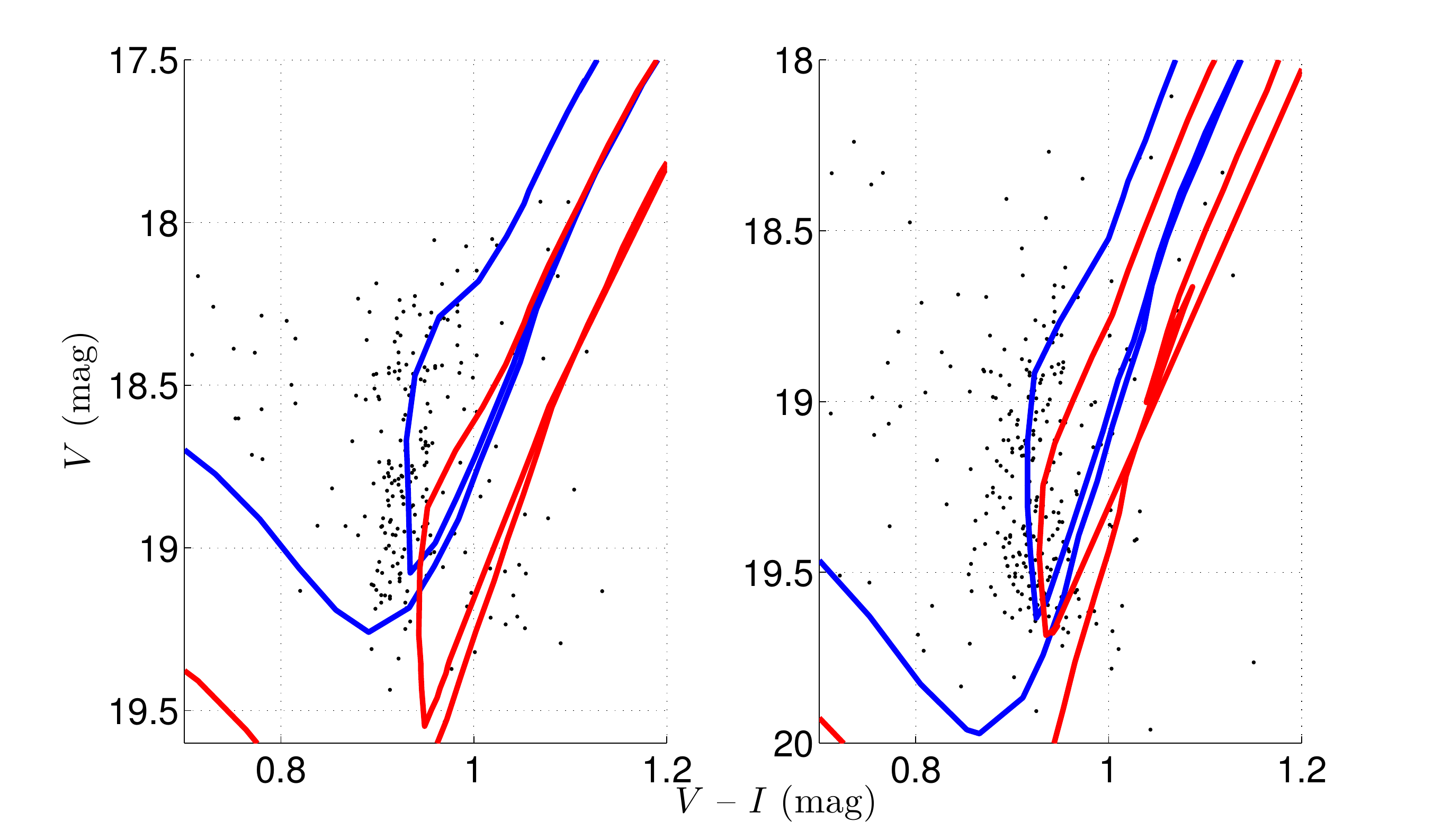}
\caption{Enlarged CMD regions centered on the clusters' RCs. The
  youngest (blue) and oldest (red) isochrones adopted are
  indicated. (left) NGC 1831. (right) NGC 1868.}
\label{fig14}
\end{figure}

The compactness of the RC offers another tight constraint in the
context of the age-dispersion model. Any age dispersion would also
affect the size of the observed RC. Therefore, we further explored the
quality of our isochrone fits to the RC region. In Fig. \ref{fig14},
we display the corresponding, zoomed-in regions centered on the
RC. For NGC 1831 (left-hand panel), we found that a dispersed
star-formation history would lead to a relatively large spread in the
RC in the magnitude direction. However, the compact size (in CMD
space) of the observed RC further constrains the cluster's maximum age
dispersion to fall within the range $8.74 \le \log(t \mbox{ yr}^{-1})
\le 8.81$. Clearly, the oldest isochrones we have adopted based on the
NGC 1831 MS spread, $\log(t \mbox{ yr}^{-1}) = 8.92$, cannot fit the
compact, observed RC. However, based only on our analysis of the
extent of the MS TO region, Fig. \ref{fig13} indicates that a
significant fraction of the cluster's stellar population may have ages
$\log(t \mbox{ yr}^{-1}) \geq 8.81$. This apparent conflict renders
applicability of the age-dispersion model for NGC 1831 uncertain. The
right-hand panel of Fig. \ref{fig14}, which shows the equivalent
analysis for the NGC 1868 RC, implies that the age-dispersion model
provides a more consistent picture for NGC 1868.

In summary, if we adopt the age-dispersion model, it is not difficult
to reproduce the observed MS TOs in detail. We only need to carefully
adjust the clusters' internal age distribution. However, the observed
size of the RC in CMD space imposes an additional constraint on the
most appropriate age range, so that the relevant age (age range) of
our sample clusters is not a completely free parameter.

\section{Fast Stellar Rotation}

The main effect of rapid stellar rotation in the context of our
CMD-based analysis can be summarized as follows:

\begin{enumerate}
\item The centrifugal force associated with rapid rotation only
  partially balances a star's self-gravity, which thus causes the star
  to expand. The expansion reduces the star's temperature, which in
  turn renders its observed color redder than that of its non-rotating
  counterparts.
\item The reduced gravity also decreases the star's core pressure,
  which reduces the amount of nuclear energy released \citep{Fauk68},
  and thus the observed rapidly rotating star becomes relatively
  fainter than the equivalent non-rotating stars.
\item Rotation affects the sizes of stellar convective
  cores. Increased convective core sizes will prolong the stellar MS
  lifetime \citep{Maed00,Yang13}.
\end{enumerate}

\cite{Bast09} first suggested that extended TOs could be explained by
rapid stellar rotation, while \cite{Yang13} suggested that the
broadening of the TOs depends on a cluster's intrinsic age. However,
neither of these authors directly compared their results with
observations. \cite{Li12} compared simulated CMDs that included
rapidly rotating stars with the observed CMDs of NGC 1846 and NGC 1987
\citep{Mack08,Milo09}. They claimed that fast rotation is necessary to
reproduce the observations. However, their comparison was only based
on the perceived similarities of their model CMDs to observational
counterparts (they do not directly compare their model CMDs with
observed CMDs; they merely refer the reader to published observational
CMDs). In this paper, we directly and quantitatively compare simulated
CMDs to our observational data.

In our simulations, we first assume a single age for the observed
stars. We opted to select the `youngest age' that we used in the
age-dispersion model, because fast rotation results in redder and
fainter stars. Following \cite{Bast09}, stars less massive than $1.2
M_{\odot}$ are assumed to lack significant rotation because of
magnetic braking \citep{Scha62,Mest87}. Stars which have evolved off
the MS do not rotate rapidly either, even if they originally rotated
rapidly, because the expansion associated with the SGB and RGB
evolutionary phases slows down their rotation velocity because of
conservation of angular momentum. Since fast rotation only affects MS
stars that are more massive than $1.2 M_{\odot}$, this effect will not
broaden either the faint section of the MS or the compact RC.

In the age-dispersion model, the age distribution determines the
observed pseudo-color distribution. For the fast-rotation model, the
distribution of the stellar rotation velocities is the operational
parameter. However, this distribution cannot be taken at will,
although the exact distribution of rotation velocities is as yet
unknown {\it for cluster stars}.

\begin{figure}[ht!]
\plotone{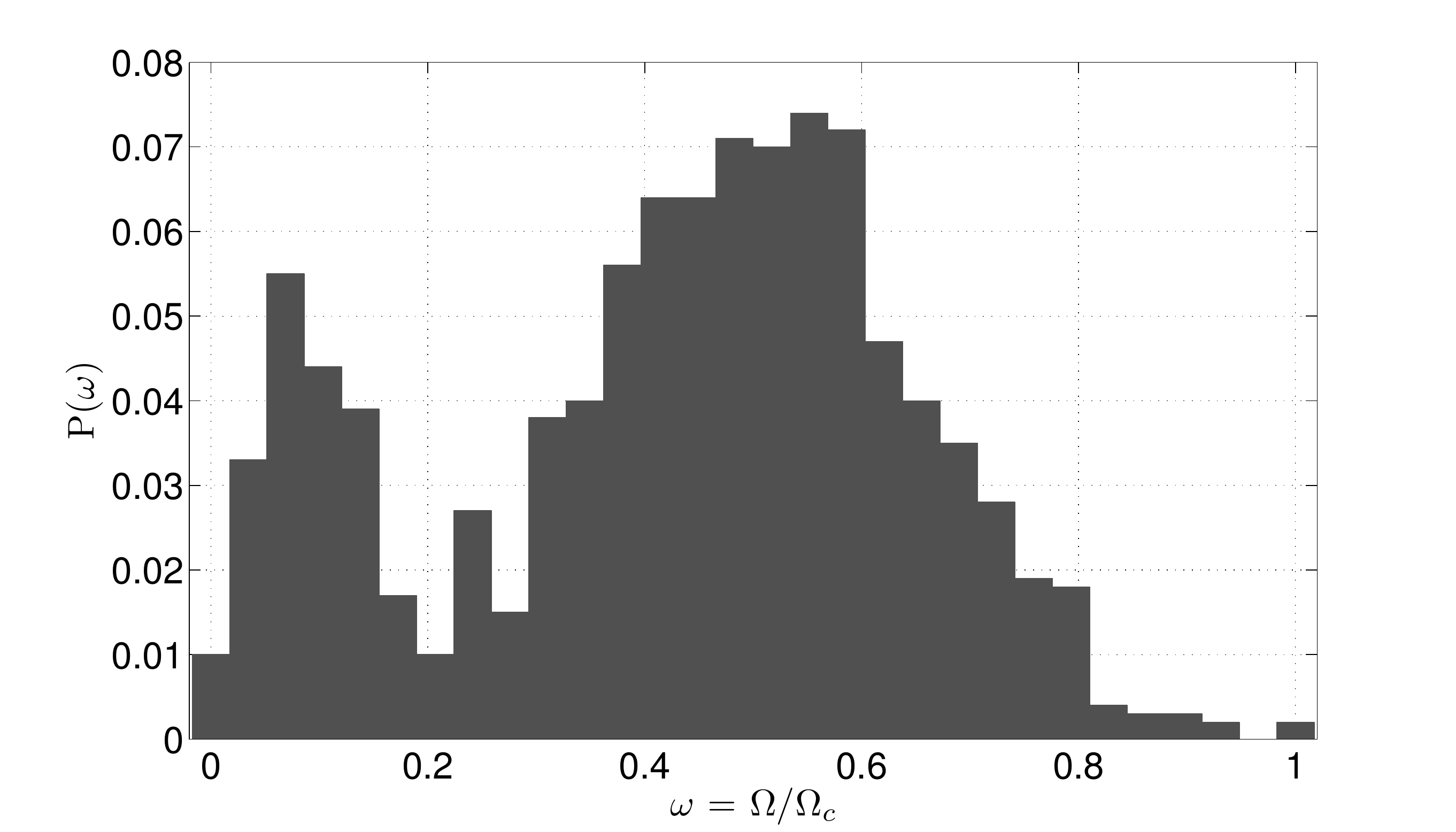}
\caption{$\omega$ (fraction of the critical break-up rate)
  distribution of rotating stars, with double Gaussian peaks at 0.10
  and 0.50, and standard deviations of 0.05 and 0.15, respectively.}
\label{fig15}
\end{figure}

\cite{Roye07} analyzed the rotation velocities of 1100 late-B to
late-F-type stars. They found that the rotation velocities of
F0--F2-type stars typically peaks at both $\omega$ $\sim 0.1$ and 0.5
(their fig. 10), where $\omega = \Omega/\Omega_{\rm c}$ is the
fraction of the critical break-up rotation rate. The peak at $\omega =
0.1$ they find is sharper than the peak at $\omega = 0.5$. The latter
peak also exhibits a broadening towards higher rotation
rates. \cite{Bast09} artificially assumed a Gaussian distribution of
rotation velocities with a peak at $\omega = 0.4$. They excluded stars
with $\omega \geq 0.7$, because above this value the assumptions in
the (one-dimensional) stellar evolution code start to break down. In
this paper, we adopt the $\omega$ distribution of \cite{Roye07}. We do
not impose any threshold on $\omega$, except that we limit $\omega$ to
be less than unity (otherwise stars will become unstable and are
disrupted by the centrifugal force), but we note that the fraction of
stars with $\omega \ge 0.7$ is only 2.5\% (74 out of 3500 stars). We
assume that the rotation velocities peak at both $\omega = 0.10$ and
0.50, and that these peaks host 20\% and 80\% in number of the stellar
sample, respectively. The standard deviations corresponding to both
peaks are 0.05 and 0.15, respectively. Figure \ref{fig15} shows the
probability distribution adopted for $\omega$. Our adopted $\omega$
distribution is similar to that of F-type stars \citep{Roye07}.

Stellar rotation velocities are mass-dependent. We assume that the
rotation velocities increase linearly for stars with masses from $1.2
M_{\odot}$ to $1.65 M_{\odot}$ \citep{Bast09}. Stars more massive than
$1.65 M_{\odot}$ are assumed to be fast rotators, with $\omega$
distributed according to Fig. \ref{fig15}. Stars with masses between
$1.2 M_{\odot}$ and $1.65 M_{\odot}$ follow the same $\omega$
distribution, adjusted by a factor between 0 (for a stellar mass of
$1.2 M_{\odot}$) and 1 (for $1.65 M_{\odot}$). Stars less massive than
1.2 $M_{\odot}$ are assumed to be non-rotating.

\begin{figure}[ht!]
\plotone{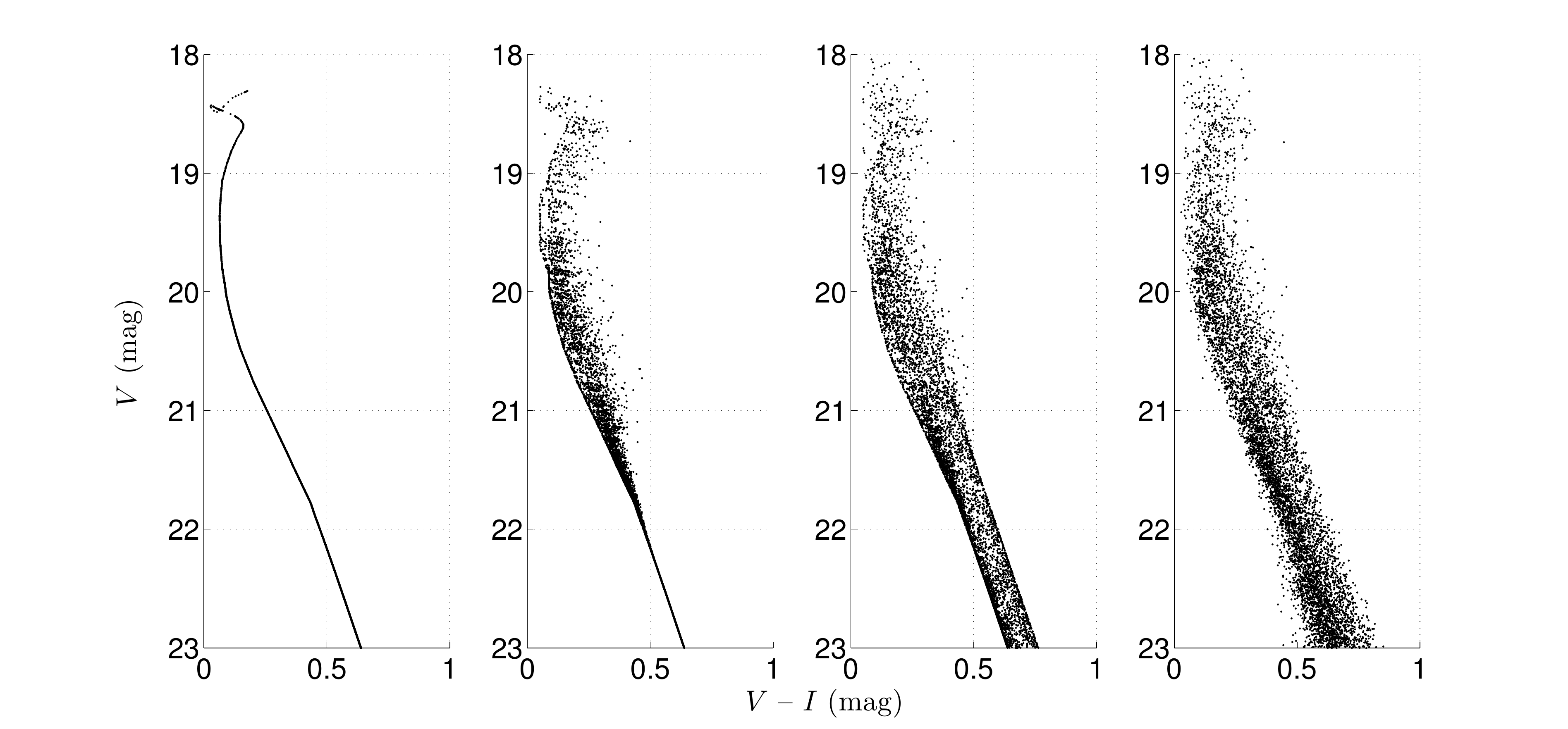}
\caption{Steps to generate our simulated NGC 1831 CMD. From left to
  right: (1) We generate stars that exactly match the parameters given
  by the adopted isochrone. (2) For stars more massive than $1.2
  M_{\odot}$, we randomly assign rotation velocities, based on the
  $\omega$ distribution of Fig. \ref{fig15}. (3) We assign `binary
  status' to 70\% of the artificial stars and adjust their photometry
  based on the adopted binary properties. (4) We adopt the appropriate
  photometric uncertainties according to Eq. (1).}
\label{fig16}
\end{figure}

Following \cite{Bast09}, the stellar temperatures and luminosities
modified by rotation are given by
\begin{equation}
  T_{\rm eff}(\omega)/T_{\rm eff}(0) = 1 - a\omega^2, 
\end{equation}
and
\begin{equation}
  L_{\rm eff}(\omega)/L_{\rm eff}(0) = 1 - b\omega^2, 
\end{equation}
where $a$ and $b$ cover the ranges 0.17--0.19 and 0.03--0.07,
respectively, during MS evolution. Since it is not fully understood
how rotation affects stars at the masses of interest, these predicted
range are uncertain. Therefore, we adopt the average values, $a =
0.18$ and $b = 0.05$, respectively, and emphasize caution in their use
at face value. First, we generate a population of stars that exactly
match the isochrone. For stars more massive than $1.2M_{\odot}$, we
randomly adopt a rotation velocity drawn from the $\omega$
distribution assumed (Fig. \ref{fig15}). We subsequently calculate the
resulting stellar temperature and luminosity, based on our
  adopted theoretical isochrone set, and convert these values to a
color and magnitude based on the theoretical isochrone set. We also
select some of the stars thus generated as the primary components of
unresolved binary systems. By adopting a flat mass-ratio distribution,
we correct their magnitudes and colors according to Eq. (1). Finally,
we adopt the appropriate photometric uncertainties to mimic the
observed CMD. Figure \ref{fig16} illustrates our procedure for NGC
1831.

\begin{figure}[ht!]
\plotone{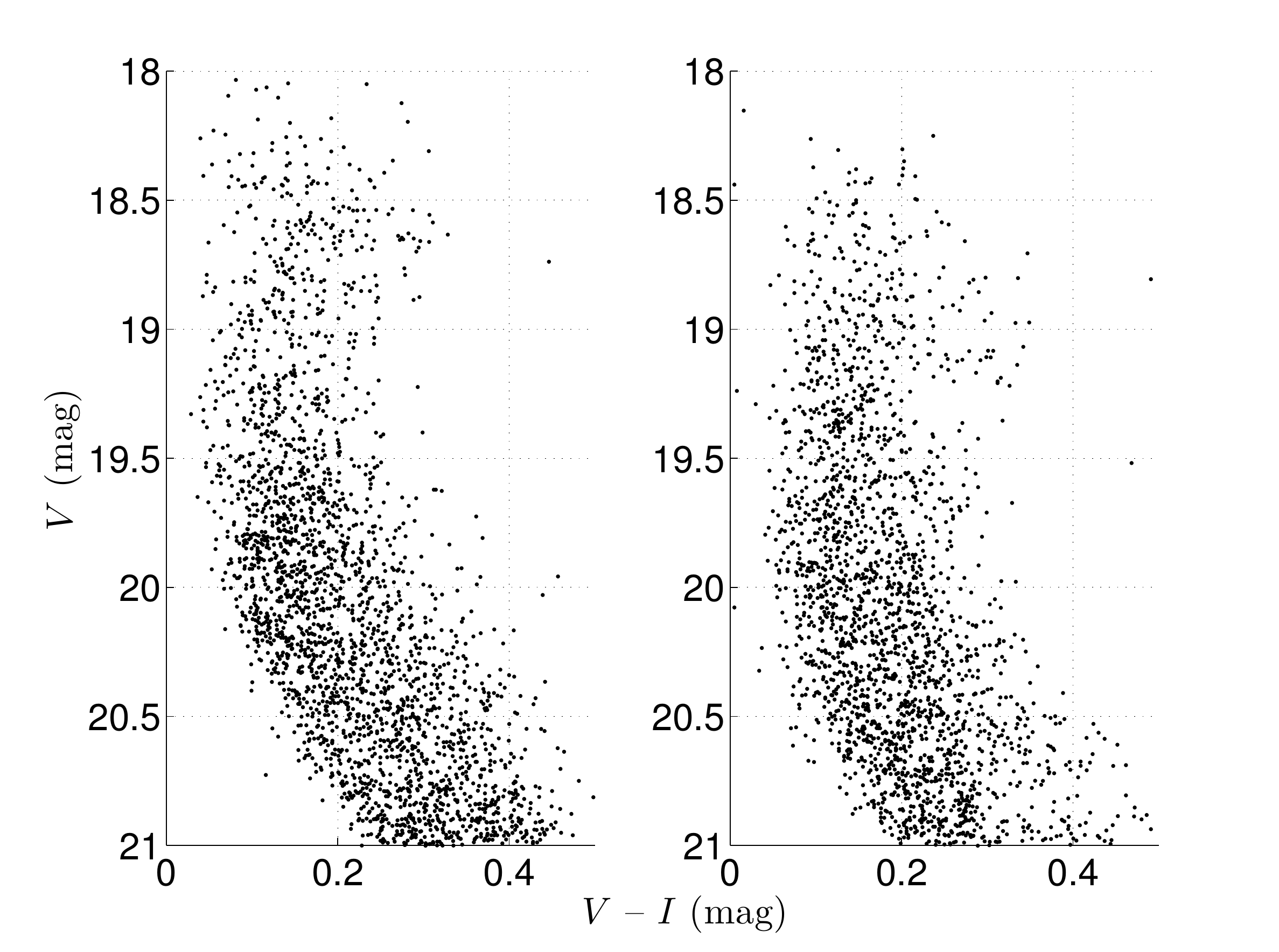}
\caption{Comparison of (left) the simulated and (right) the observed
  NGC 1831 CMDs, based on the fast-rotation model.}
\label{fig17}
\end{figure}

\begin{figure}[ht!]
\plotone{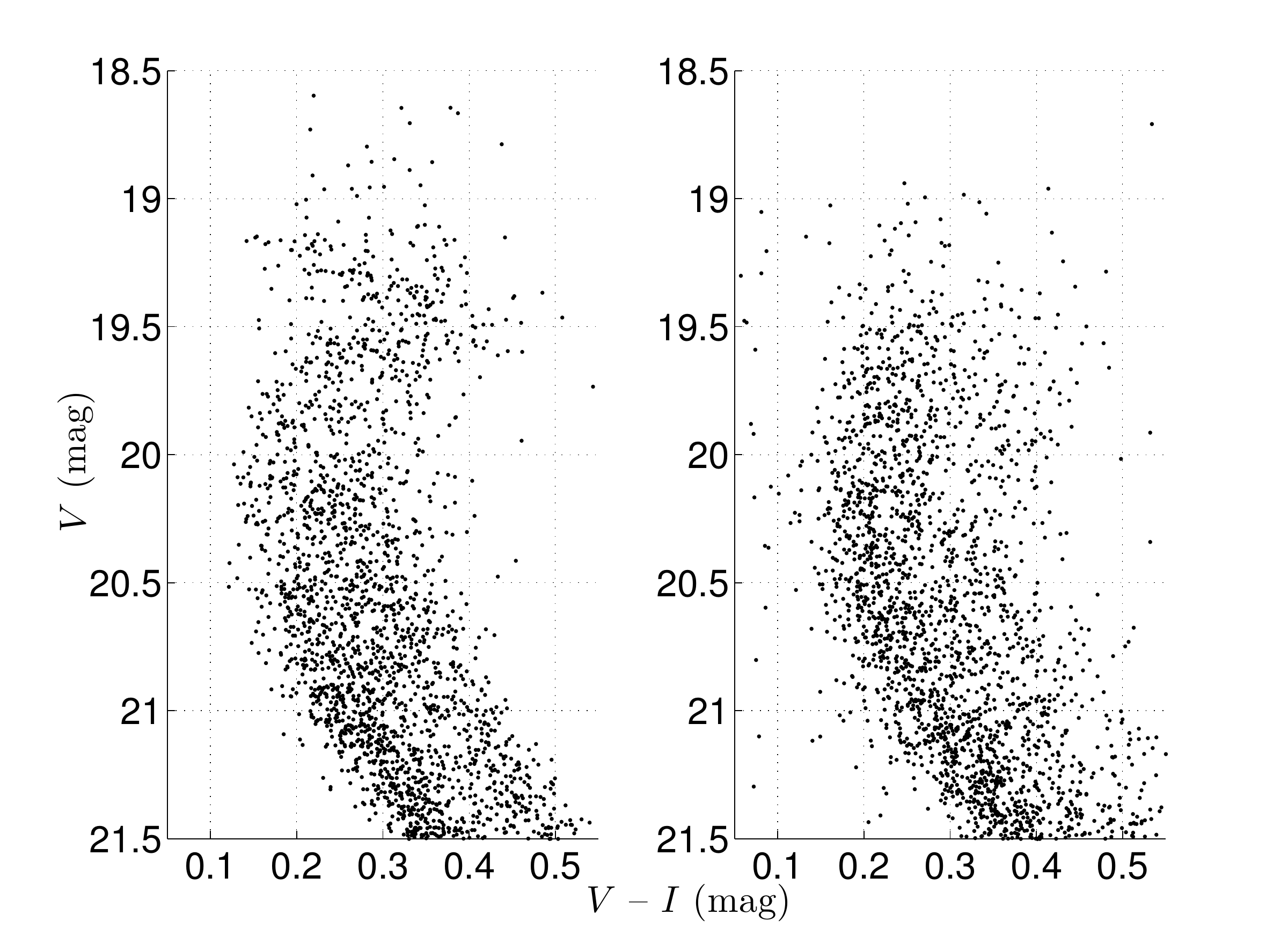}
\caption{As Fig. \ref{fig17}, but for NGC 1868.}
\label{fig18}
\end{figure}

\begin{figure}[ht!]
\plotone{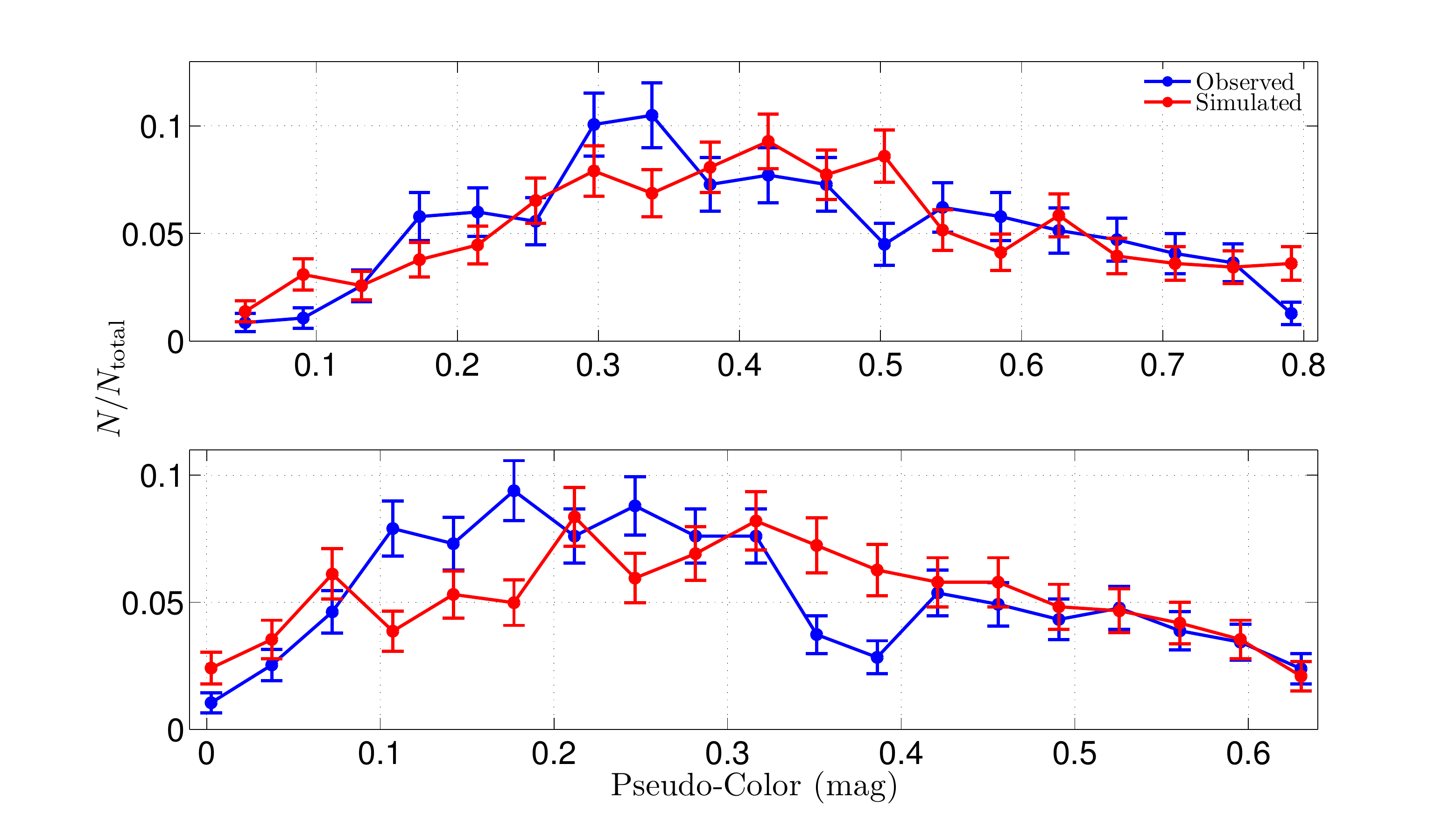}
\caption{Pseudo-color distributions of (red) the simulated MS TO stars
  and (blue) the observed MS TO stars for (top) NGC 1831 and (bottom)
  NGC 1868.}
\label{fig19}
\end{figure}

In Figs \ref{fig17} and \ref{fig18}, we show the simulated clusters'
CMDs, as well as their observed counterparts. Following our approach
as applied to the age-dispersion model, we again explore the
pseudo-color distributions for both simulated CMDs (cf. Figs
\ref{fig7} and \ref{fig6}). Figure \ref{fig19} displays the
comparison. Again, we run K--S tests on our results to quantify the
degree of similarity. The null hypothesis is rejected, which implies
that the fast-rotation model indeed succeeds in reproducing the
observations adequately. Again, the K--S test yields a
  probability of 74\% that the simulated distributions of both NGC 1831
  and NGC 1868 have the same underlying distribution as the
  observations. This implies that the fast-rotation model is able to
appropriately reproduce the observational data to the same extent as
the age-dispersion model. This is encouraging, because contrary to the
age-dispersion model -- where we needed to adjust the age
distributions to reproduce the observations -- there is no need adopt
a special $\omega$ distribution.

\label{disc.sec}
    
\section{Discussion and Conclusions}

We have systematically analyzed the extended TOs associated with the
intermediate-age LMC clusters NGC 1831 and NGC 1868. We rule out
differences in helium abundances as the cause of the extended MS TOs,
because of the observed narrow MS and compact RC features. Based on
the observational data, we apply a carefully considered set of
criteria to select possible MS TO stars for further investigation of
their pseudo-color distributions. Using the age-dispersion and
fast-rotation models, we generate artificial star clusters to explore
which of these models best match the observations.

If we only consider the stars in the extended MS TO region, both
models perform equally well. For NGC 1831, the adopted age ranges from
$\log(t \mbox{ yr}^{-1}) = 8.74$ to 8.92, corresponding to an age
dispersion of $\sim 280$ Myr. We found that if we adopt a dominant age
of (i.e., a peak in the age distribution at) $\log(t \mbox{ yr}^{-1})
= 8.80$, with gradually decreasing wings on either side of the peak,
we obtain the highest degree of similarity of the model versus
observed pseudo-color distributions. For NGC 1868, it is harder to
find a best-fitting model, although we still manage to find an
appropriate representation: the age-dispersion model indicates an age
range from $\log(t \mbox{ yr}^{-1}) = 8.93$ to 9.07 ($\sim 320$ Myr).
The corresponding best-fitting age distribution decreases smoothly
from $\log(t \mbox{ yr}^{-1}) = 8.98$ to 9.07. Both clusters exhibit a
similar age difference, which is also consistent with the age ranges
inferred for most intermediate-age clusters in the LMC. The origin of
any age dispersion within such intermediate-age clusters is still
unclear, however.

On the other hand, if we take the compactness (in CMD space) of the
clusters' RCs into consideration, we can set further constraints on
the likely maximum age ranges appropriate for either cluster. For NGC
1868, the age-dispersion model is still consistent with the results of
a model including rapid rotation, but if the fast-rotation model for
NGC 1831 is correct, the corresponding age dispersion would be much
smaller---$\log( t \mbox{ yr}^{-1}) = 8.74$ to 8.81---than that
inferred from the age-dispersion model.

In the context of the fast-rotation model, the $\omega$ distributions
used in previous studies have thus far remained unconstrained
\citep{Bast09,Li12,Yang13}, in essence because the $\omega$
distribution appropriate for cluster stars is unknown. Here, we adopt
the $\omega$ distribution derived by \cite{Roye07} for F-type field
stars. Based upon this strong constraint, the fast-rotation model can
be used to satisfactorily reproduce the observations of both of our
sample clusters. K--S tests imply a significant similarity between the
model and observed CMDs (pseudo-color distributions). In particular
for NGC 1831, we argue that the fast-rotation model provides a
preferred description of the data; validation of the age-dispersion
model would imply that the age range of cluster stars must be reduced
from 280 to 100 Myr, which is not borne out by our analysis based on
the extent of the cluster's TO region. For NGC 1868, both models work
equally well.

If the validity of the age-dispersion model is confirmed, this may
indicate that some dramatic events have occurred on a global scale
across the LMC. Intermediate-age clusters that are located in close
proximity of each other should have experienced similar star-formation
histories. Both NGC 1831 and NGC 1868 are located on the same side of
LMC, close together in the same (fourth) quadrant, with a projected
distance of less than 1.25 kpc. However, the age ranges inferred for
NGC 1831 and NGC 1868 do not overlap, and even their age distributions
appear to differ significantly. In this paper, we have found that the
fast-rotation model satisfactorily reproduces the
observations. However, it is unclear whether the $\omega$ distribution
adopted from \cite{Roye07} is realistic for cluster stars. This
provides an incentive for future work in this area. We also note that
even though most observed intermediate-age LMC clusters display
extended TOs,\footnote{Of the 16 intermediate-age LMC clusters
  analyzed by \cite{Milo09}, $70 \pm 25$\% ($\sim 11$ clusters)
  exhibit CMD features that are inconsistent with those of simple
  stellar populations. Of these, three clusters show multiple distinct
  TOs, while the remaining eight display CMDs that are similar in
  nature to those of NGC 1831 and NGC 1868 discussed here. Note that
  our two sample clusters were not included in the study of
  \cite{Milo09}.} a small fraction display multiple distinct,
well-defined TOs, e.g., NGC 1846 \citep{Mack07,Milo09}, NGC 1806, and
NGC 1751 \citep{Milo09}. Clusters exhibiting such distinct MS TOs most
likely include stars of different (distinct) ages, since the
smoothly-varying fast-rotation model cannot account for such features.
  
\section*{Acknowledgements}

We thank Zhongmu Li, Song Huang, and Selma de Mink for offering useful
suggestions and assistance. We are grateful for support from the
National Natural Science Foundation of China through grants 11073001,
10973015, and 11373010.

\end{document}